\documentclass[structabstract]{aa}  
\usepackage{graphicx}
\usepackage{txfonts}
\usepackage{natbib}
\bibpunct{(}{)}{;}{a}{}{,} 

\begin{document}
\title{Pulsating low-mass white dwarfs in the frame of new evolutionary 
sequences} 
\subtitle{II. Nonadiabatic analysis}
\author{A. H. C\'orsico\inst{1,2} \and L. G. Althaus\inst{1,2}}
\institute{$^{1}$ Grupo  de Evoluci\'on  Estelar y  Pulsaciones,  Facultad de
           Ciencias Astron\'omicas  y Geof\'{\i}sicas, Universidad  Nacional de
           La Plata, Paseo del Bosque s/n, (1900) La Plata, Argentina\\  
           $^{2}$Instituto de Astrof\'{\i}sica La Plata, CONICET-UNLP, Paseo 
           del Bosque s/n, (1900) La Plata, Argentina\\
            \email{acorsico,althaus@fcaglp.unlp.edu.ar}     
           }
\date{Received ; accepted }

\abstract{Low-mass  ($M_{\star}/M_{\sun}  \lesssim 0.45$) white
  dwarfs,   including the  so  called extremely low-mass white dwarfs
  (ELM, $M_{\star}/M_{\sun } \lesssim  0.18-0.20$), are being
  currently discovered  in  the  field  of  our  Galaxy  through
  dedicated   photometric  surveys.    The  fact that some of them
  pulsate opens the unparalleled chance for sounding  their interiors.}    
  {We  present  a  detailed nonadiabatic
  pulsational  analysis of such   stars employing full evolutionary
  sequences of low-mass He-core white dwarf models derived  from
  binary star evolution computations. The  main aim  of this study  is
  to  provide a detailed description of the pulsation stability
  properties of variable low-mass  white dwarfs during the terminal
  cooling branch .}{Our nonadiabatic
  pulsation analysis is based  on   a new set of He-core white-dwarf
  models with  masses ranging   from  $0.1554$ to $0.4352  M_{\sun}$
  derived  by computing the  non-conservative evolution  of a  binary
  system consisting of  an initially $1 M_{\sun}$  ZAMS star and a
  $1.4 M_{\sun}$ neutron star.   We have computed nonadiabatic radial
  ($\ell= 0$) and nonradial ($\ell= 1, 2$) $g$ and $p$ modes to assess
  the dependence of the  pulsational stability properties of these
  objects  with stellar parameters such  as the stellar mass, the
  effective  temperature, and the convective efficiency.}  {We found
  that a dense spectrum of unstable radial modes and nonradial $g$ and
  $p$ modes  are driven by the $\kappa-\gamma$ mechanism due to the
  partial ionization of H in the stellar envelope, in addition to
  low-order unstable $g$ modes characterized by short pulsation
  periods which  are significantly excited by H burning via the
  $\varepsilon$  mechanism of mode driving. In all the cases, the
  characteristic  times required for the modes to reach  amplitudes
  large  enough as  to be observable (the $e$-folding times) are
  always shorter than  cooling timescales. We explore the dependence
  of the ranges of  unstable mode periods (the longest and shortest
  excited periods)  with the effective temperature, the stellar mass,
  the convective  efficiency, and the harmonic degree of the modes. We
  also compare our  theoretical predictions with the excited modes
  observed in the seven  known variable low-mass white dwarfs (ELMVs),
  and found an  excellent agreement.}{} 

\titlerunning{Nonadiabatic pulsations of low-mass white dwarfs}
   \maketitle
%

\section{Introduction}
\label{introduction}

White dwarf  (WD) stars constitute the  last stage in the  life of the
majority ($\sim  97 \%$) of  stars populating the  Universe, including
our  Sun \citep{2008ARA&A..46..157W,2008PASP..120.1043F,review}.  Most
of WDs show H rich atmospheres, defining the spectral class of DA WDs.
The mass  distribution of  DA WDs peaks  at $\sim 0.59  M_{\sun}$, and
exhibits      also     high-mass      and      low-mass     components
\citep{2007MNRAS.375.1315K,2011ApJ...730..128T,2013ApJS..204....5K,2015MNRAS.446.4078K}.
Low-mass  ($M_{\star}/M_{\sun}  \lesssim  0.45$)  WDs are the result
of strong mass-loss episodes in interacting binary systems during  the
red giant branch stage of low-mass stars before the onset of  helium
flash \citep[see, for recent
  works,][]{2013A&A...557A..19A,2014A&A...571A..45I}.   Since  the
ignition of  He is avoided,  they probably harbor cores of He, at
variance with average mass   WDs  which are expected to have  cores
made  of C and  O. In particular,  interacting binary  evolution  is
the  most likely  origin  for  the extremely low-mass (ELM) WDs,
which have masses below $\sim 0.18-0.20 M_{\sun}$.
State-of-the-art     evolutionary    computations    of
\citet{2013A&A...557A..19A}   \citep[see  also][]{2001MNRAS.323..471A,
  2007MNRAS.382..779P,2014A&A...571A..45I}, predict  that ELM WDs must
be characterized  by very  thick H envelopes  which should be  able to
sustain  residual H nuclear  burning via  $pp$-chain, thus  leading to
very long evolutionary timescales  ($\sim 10^{9}$ yrs). In comparison,
low-mass WDs  with $M_{\star} \gtrsim 0.18-0.20  M_{\sun}$ should have
cooling timescales of the order  of $\sim 10^{7}$ yrs. This is because
their   progenitors    experience   multiple   diffusion-induced   CNO
thermonuclear  flashes  that engulf  most  of  the  H content  of  the
envelope. As a result, the remnants enter their final cooling tracks
with a  very thin H  envelope, which  is unable  to sustain  stable
nuclear burning while cools.

\begin{figure} 
\begin{center}
\includegraphics[clip,width=8 cm]{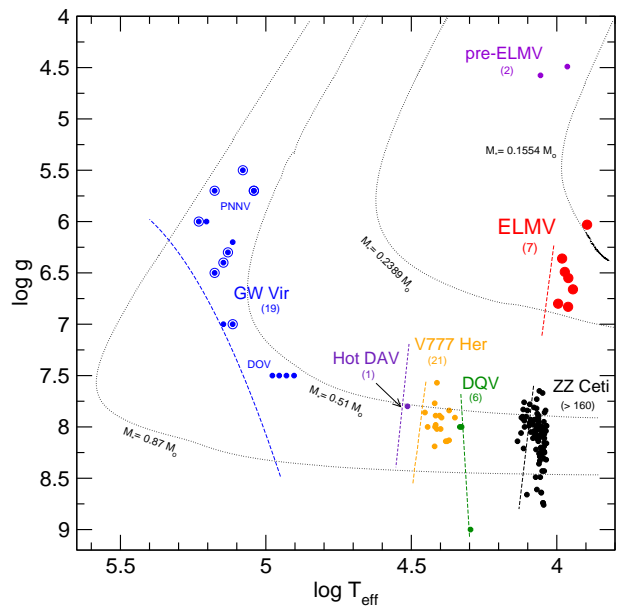} 
\caption{The location of the known ELMVs (big red circles) along with the 
  other several classes of pulsating WD stars (dots of different colors) 
  in the   $\log T_{\rm eff} - \log g$ plane.  
  In parenthesis we include the number of known members of
  each class.  Two post-VLTP (Very Late Thermal Pulse) 
  evolutionary  tracks  for H-deficient WDs and two 
  evolutionary tracks for low-mass He-core WDs  are  plotted   for reference.  
  Also   shown  is  the  theoretical
  blue   edge  of  the instability strip for  the GW Vir stars
  \citep{2006A&A...458..259C}, the hot DAV stars, 
\citep{2013EAS....63..185S}, V777 Her stars \citep{2009JPhCS.172a2075C}, 
  the  DQV stars \citep{2009A&A...506..835C}, the ZZ Ceti  stars
  \citep{2008PASP..120.1043F}, and the pulsating  low-mass  WDs
  \citep{2013MNRAS.436.3573H}. }
\label{figure_01} 
\end{center}
\end{figure}

Many low-mass WDs,  including ELM WDs, are being currently detected
through the ELM      survey     and     the      SPY     and      WASP
surveys \citep[see][]{2009A&A...505..441K,2010ApJ...723.1072B,
  2012ApJ...744..142B,2011MNRAS.418.1156M,2011ApJ...727....3K,
  2012ApJ...751..141K,2013ApJ...769...66B,2014ApJ...794...35G,2015MNRAS.446L..26K}.
Interest in these stars has strongly increased following the discovery
that  some of  them pulsate  with periods  compatible  with high-order
nonradial         $g$         modes        \citep{2012ApJ...750L..28H,
  2013ApJ...765..102H,2013MNRAS.436.3573H,2015MNRAS.446L..26K,
  2015ASPC..493..217B},  providing  the   chance  for  sounding  their
interiors by  employing asteroseismology. At the time  of writing this
paper,   seven  pulsating   ELM   WDs  (hereafter   ELMVs\footnote{For
  simplicity, here  and throughout  the paper we  will refer  to the
  pulsating  low-mass  WDs  as  ELMVs,  even in  the  cases  in  which
  $M_{\star}   \gtrsim  0.18-0.20  M_{\sun}$.})   are  known.   It  is
interesting to put the  ELMVs  in the  context  of the  other classes
of pulsating WDs.   In Fig.  \ref{figure_01} we show  the location
of the  ELMVs (big red  circles),  along  with the  several families
of pulsating WDs known hitherto.  The  ELMV instability domain can be
seen as an extension of the ZZ Ceti instability strip towards low
effective temperatures and  gravities. The variables ZZ Ceti  or DAVs
(pulsating DA  WDs,  with  almost  pure  H atmospheres)  are  the
most  numerous ones. The other classes comprise  the DQVs (atmospheres
rich in He and C), the  variables V777 Her or  DBVs (atmospheres
almost  pure in He), the  Hot DAVs  (H-rich  atmospheres),  and  the
variables  GW  Vir (atmospheres  dominated by C,  O, and  He) that
include the  DOVs and PNNVs  objects.  To  this  list,  we  have to
add  the  newly  discovered pre-ELMVs
\citep{max13,2014MNRAS.444..208M}, the  probable precursors of ELMV
stars. Note that the effective temperatures of ELMVs are found to be
between $\sim 10\,000$ K and $\sim 7800$ K, and so, they are the
coolest pulsating WDs known up to date (see Fig. \ref{figure_01})

The classification of ELMV stars as a new, separate class of pulsating
WDs   is   a  matter   of   debate.   On   one  hand,   some   authors
\citep[e.g.][]{2013ApJ...762...57V} consider  the ELMVs as  genuine ZZ
Ceti stars but  with very low masses. This conception  is based on the
fact  that   for  both  kinds   of  objects  (which  share   the  same
spectroscopic  classification  as  DA  objects),  the  pulsations  are
excited by  the same driving    $\kappa-\gamma$ mechanism associated
with the H partial ionization zone.   However, there are  significant
differences between both  types of  stars.   From the  point  of view
of  their origin  and formation, the low-mass WD stars (including ELM
WDs) seem to come from interacting binary evolution and should harbor
cores made of unprocessed He. This is in contrast to the case of 
average-mass ZZ Cetis, that according to the standard evolutionary 
theory, are the result of single-star evolution and must have  
cores made  of C  and O.  In addition,  the fact  that  so 
many constant (non variable) low-mass WDs coexist with ELMV 
WDs in the same domain of $T_{\rm   eff}$ and 
$\log g$\footnote{That is,  the domain of instability
  seems  not to be  ``pure'' \citep{2013MNRAS.436.3573H}}, may be
indicating substantially  different internal structures and so, quite
distinct evolutionary origins.  This  is in contrast to the well
documented purity of  the ZZ Ceti instability strip,  which indicates
that  all the DA  WDs crossing  the effective temperature interval
$12\,500\ {\rm K} \gtrsim  T_{\rm  eff} \gtrsim 10\,700$ K do pulsate.
Another distinctive  feature of  ELMVs  is the length of their
pulsation periods,  that largely exceed $\sim 1200$ s and reach  up to
$\sim 6300$  s, much longer  than the periods found in ZZ Ceti stars
($100\ {\rm s} \lesssim \Pi \lesssim 1200$ s).  Indeed, the   period
at   $\Pi=   6235$   s  detected   in   the  ELMV   SDSS
J222859.93+362359.6 \citep{2013MNRAS.436.3573H}  is the longest period
ever measured in a pulsating WD star.

On    the    theoretical    side,    the   pulsation    analysis    by
\citet{2014A&A...569A.106C}   constitutes  at present  the   most
detailed   and exhaustive  investigation  of  the  adiabatic
properties  of  low-mass WDs.  The background  equilibrium  models
employed by  those authors were extracted from the complete set of
evolutionary sequences of  low-mass He-core WD models  presented in
\citet{2013A&A...557A..19A}. The results        of
\citet{2014A&A...569A.106C}        \citep[see also the pioneering
  works of][]{2010ApJ...718..441S, 2012A&A...547A..96C} indicate  that
$g$ modes  in ELMVs  are restricted  mainly  to the  core regions  and
$p$ modes to the envelope, providing  the chance to constrain both the
core   and   envelope   chemical   structure  of   these   stars   via
asteroseismology.  On  the other  hand,  nonadiabatic studies
\citep{2012A&A...547A..96C, 2013ApJ...762...57V}  predict   that  many
unstable  $g$ and $p$ modes are excited by  the same  partial
ionization mechanism at  work in  ZZ Ceti stars,  roughly at the
right effective temperatures and the correct range of the periods
observed in ELMVs.  In this paper, our second work of the series on
this topic, we perform a  thorough   stability  analysis  on  the
set  of  state-of-the-art evolutionary models  of
\citet{2013A&A...557A..19A}.   Preliminary results of this analysis
were presented in the work of  \citet{2014ApJ...793L..17C},  focused
mainly on  the role  that stable  H burning  has in destabilizing
low-order $g$ modes of  ELM WD  models. The  results of that  paper
constitute  the  first theoretical  evidence of  pulsation modes
excited  by the $\varepsilon$ mechanism in cool  WD stars. Here, we
extend that analysis by assessing the vibrational  stability of radial
($\ell= 0$) and nonradial ($\ell= 1, 2$) $p$ and $g$  modes for the
complete set of 14 evolutionary sequences of
\citet{2013A&A...557A..19A} with masses in the  range $0.1554-0.4352
M_{\sun}$, considering both  the  $\kappa-\gamma$  and
$\varepsilon$ mechanisms of mode driving,  and including different
prescriptions of the MLT theory of convection.

The paper is organized as  follows.  In Sect.  \ref{models} we briefly
describe  our  numerical  tools   and  the  main  ingredients  of  the
evolutionary sequences  we employ to  assess the nonadiabatic
pulsation  properties of low-mass He-core WDs.   In Sect.
\ref{stability} we present in detail our pulsation  results.
Sect. \ref{obs} is devoted to compare the predictions of our
nonadiabatic models with the ranges of excited periods in the observed 
stars. Finally,    in Sect. \ref{conclusions} we summarize  the main findings
of the paper.

\section{Computational tools and stellar models}
\label{models}

\subsection{Evolutionary code}
\label{evol_code}

The  evolutionary WD models  employed  in our  pulsational analysis
were generated with  the {\tt LPCODE} evolutionary code,  which
produces complete and detailed  WD models incorporating  very updated
physical ingredients. In addition, {\tt LPCODE} computes in detail the full
evolutionary stages leading to the WD formation, allowing to study the
WD evolution in a consistent way with the expectations of the
evolutionary  history of progenitors. While detailed information
about {\tt LPCODE} can be found in
\citet{2005A&A...435..631A,2009A&A...502..207A,2013A&A...557A..19A}
and references therein, we list below only  those   ingredients
employed  which  are  relevant for  our  analysis  of  low-mass,
He-core WD stars:

\begin{itemize}

\item  [-] The standard  Mixing Length  Theory (MLT)  for convection
  in the versions ML1, ML2 and  ML3 is used. The ML1 version, due to  
  \citet{1958ZA.....46..108B},  has $\alpha= 1$ and coefficients 
$a= 1/8, b= 1/2, c= 24$. The parameter $\alpha$ is the 
mixing length in units of the local pressure scale height, and
the  coefficients $a, b, c$ appear in the equations for the
average speed of the convective cell, the average convective flux, and 
the convective efficiency \citep[see][]{1968pss..book.....C}. The ML2 version, 
in turn, 
is due to \citet{1971A&A....12...21B}, and also have $\alpha= 1$,
but coefficients  $a= 1, b= 2, c= 16$. Finally, the ML3 version
is characterized by $\alpha= 2$ and the same coefficients  
$a, b$, and $c$ as in ML2. Physically, the main difference between these 
different prescriptions of MLT is the increasing convective efficiency 
going from ML1 to ML3 \citep[for details, see][]{1990ApJS...72..335T}. 

\item [-] Metallicity of the progenitor  stars has been assumed to be
  $Z = 0.01$.

\item [-] Radiative opacities for arbitrary metallicity in the range
  from 0 to  0.1 are from the OPAL project
  \citep{1996ApJ...464..943I}.  At  low temperatures, we  use the
  updated molecular  opacities with varying C/O ratios computed at
  Wichita State University  \citep{2005ApJ...623..585F} and presented
  by  \citet{2009A&A...508.1343W}.

\item [-] Conductive opacities are those of
  \citet{2007ApJ...661.1094C}.

\item [-] The  equation of state during the  main sequence evolution
  is that of OPAL for H- and He-rich composition.

\item   [-]   Neutrino  emission   rates   for   pair,  photo,   and
  bremsstrahlung processes have been taken from
  \citet{1996ApJS..102..411I},  and for plasma processes we included
  the treatment of  \citet{1994ApJ...425..222H}.

\item [-] For the WD  regime we have employed an updated version of
  the \citet{1979A&A....72..134M} equation of state.

\item [-] The nuclear network  takes into account 16 elements and 34
  thermonuclear  reaction  rates   for  pp-chains,  CNO  bi-cycle,  He
  burning, and C ignition.

\item [-] Time-dependent diffusion due to gravitational settling and
  chemical  and thermal diffusion  of nuclear  species has been  taken
  into account  following  the  multicomponent  gas  treatment  of
  \citet{1969fecg.book.....B}.

\item  [-]  Abundance changes  have been computed  according to
  element diffusion, nuclear reactions,  and convective mixing.  This
  detailed treatment  of abundance  changes by  different processes
  during the WD regime constitutes a key aspect in the evaluation of
  the importance of  residual nuclear burning for the  cooling of
  low-mass WDs.

\item [-]  For the  WD regime and  for effective  temperatures lower
  than $10\, 000$ K, outer boundary conditions for the evolving models
  are derived from nongrey model atmospheres
  \citep{2012A&A...546A.119R}.

\end{itemize}

\subsection{Pulsation code}
\label{pulsa_code}

We carry out our pulsation analysis of  radial ($\ell= 0$) and
nonradial ($\ell= 1, 2$) $p$ and $g$ modes     employing the
nonadiabatic  versions of the  {\tt LP-PUL} pulsation code described
in detail  in \citet{2006A&A...458..259C}. For the nonradial
computations, the  code solves  the  sixth-order complex system  of
linearized equations and  boundary conditions as given by
\citet{1989nos..book.....U}. For the case of radial modes,  {\tt
  LP-PUL} solves the fourth-order complex system  of linearized
equations and  boundary conditions  according to
\citet{1983ApJ...265..982S}, with the simplifications of
\citet{1993ApJ...404..294K}.  Our nonadiabatic  computations  rely on
the frozen-convection (FC) approximation,  in which  the  perturbation
of  the convective flux is neglected. While  this approximation is
known to give unrealistic locations of the $g$-mode red edge of the ZZ
Ceti instability strip,   it leads to satisfactory predictions for the
location of the blue edge \citep{2012A&A...539A..87V} \citep[see also]
[for an enlightening  discussion of this topic] {2013EPJWC..4305005S}.

\subsection{Model sequences}
\label{model_seq}

\begin{table}
\centering
\caption{Selected  properties  of our  He-core  WD  sequences
  (final cooling branch)   at  $T_{\rm eff}
  \approx 10\,000$  K: the stellar  mass, the mass  of H in  the outer
  envelope, the time
  it takes the  WD models to cool from $T_{\rm eff} \approx  10\, 000$ K to
  $\approx 8000$ K, and the occurrence (or not) of CNO flashes on the early 
   WD cooling branch.}
\begin{tabular}{cccc}
\hline
\hline
\noalign{\smallskip}
 $M_{\star}/M_{\sun}$  & $M_{\rm H}/M_{\star}\ [10^{-3}]$ & 
$\tau\ [{\rm Gyr}= 10^9 {\rm yr}]$ & H-flash\\
\hline
\noalign{\smallskip}
0.1554 & 25.4 & 3.13 &  No \\ 
0.1612 & 20.6 & 4.44 &  No \\
0.1650 & 18.7 & 5.53 &  No \\
0.1706 & 16.3 & 6.59 &  No \\
0.1762 & 14.5 & 7.56 &  No \\
\hline
\noalign{\smallskip}
0.1806 & 3.68 & 0.34 & Yes \\
0.1863 & 4.36 & 0.37 & Yes \\
0.1917 & 4.49 & 0.35 & Yes \\
0.2019 & 3.80 & 0.32 & Yes \\
0.2389 & 3.61 & 0.62 & Yes \\
0.2707 & 1.09 & 0.33 & Yes \\
0.3205 & 1.60 & 0.91 & Yes \\
0.3624 & 0.80 & 0.58 & Yes \\
0.4352 & 0.63 & 0.91 & No  \\
\noalign{\smallskip}
\hline
\hline
\end{tabular}
\label{table1}
\end{table}

\begin{figure} 
\begin{center}
\includegraphics[clip,width=9 cm]{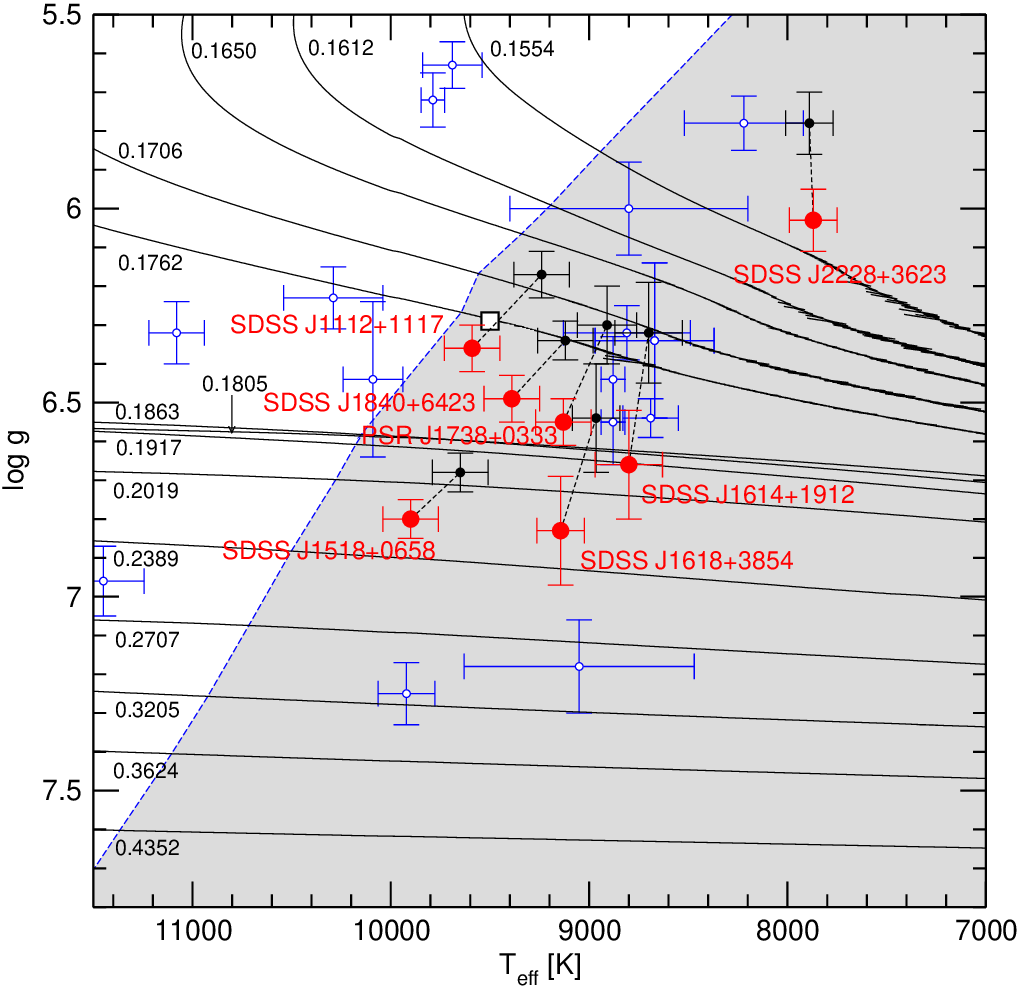} 
\caption{$T_{\rm eff} - \log g$ plane showing the low-mass He-core  WD
  evolutionary tracks (final cooling branches)  of
  \citet{2013A&A...557A..19A}. Numbers correspond to the stellar mass
  of each sequence. The locations of the seven known ELMVs
  \citep{2012ApJ...750L..28H,
    2013ApJ...765..102H,2013MNRAS.436.3573H,2015MNRAS.446L..26K,
    2015ASPC..493..217B} are marked with large red circles  ($T_{\rm
    eff}$ and $\log g$ computed with 1D model atmospheres) and small
  black circles (after 3D corrections).  Stars observed not to vary
  \citep{2012PASP..124....1S,2012ApJ...750L..28H,
    2013ApJ...765..102H,2013MNRAS.436.3573H} are depicted with hollow 
  small blue circles. The  hollow square on the 
  evolutionary track of $M_{\star}=
  0.1762 M_{\sun}$ indicates the location of the template models
  analyzed in Sect. \ref{stability}. The gray shaded  region bounded
  by the dashed blue line corresponds to the instability  domain of
  $\ell= 1$ $g$ modes according to nonadiabatic computations using ML2
  ($\alpha= 1.0$) version of the MLT theory of convection; see
  Sect. \ref{strip}.}
\label{figure_02} 
\end{center}
\end{figure}

\begin{table*}
\centering
\caption{Stellar parameters (derived using 1D and 3D model atmospheres) and observed pulsation 
properties of the seven known ELMV stars.}
\begin{tabular}{lcccccccl}
\hline
\hline
\noalign{\smallskip}
 Star & $T_{\rm eff}^{\rm (1D)}$ & $\log g^{\rm (1D)}$  & $M_{\star}^{\rm (1D)}$ & $T_{\rm eff}^{\rm (3D)}$ & 
$\log g^{\rm (3D)}$  & $M_{\star}^{\rm (3D)}$  & Period range  & Ref.\\
     &                         [K] &    [cgs] & [$M_{\sun}$] &  [K] &  [cgs] & [$M_{\sun}$]  &  [s] & \\
\noalign{\smallskip}
\hline
\noalign{\smallskip}
SDSS J222859.93$+$362359.6  & $7870\pm120$ & $6.03\pm0.08$ & $0.152$ & $7890\pm120$  & $5.78\pm0.08$ & $0.142$ & $3254-6235$ & (2) \\
SDSS J161431.28$+$191219.4  & $8800\pm170$ & $6.66\pm0.14$ & $0.192$ & $8700\pm170$  & $6.32\pm0.13$ & $0.172$ & $1184-1263$ & (2) \\
PSR  J1738$+$0333           & $9130\pm140$ & $6.55\pm0.06$ & $0.181$ & $8910\pm150$  & $6.30\pm0.10$ & $0.172$ & $1788-3057$ & (4) \\
SDSS J161831.69$+$385415.15 & $9144\pm120$ & $6.83\pm0.14$ & $0.220$ & $8965\pm120$  & $6.54\pm0.14$ & $0.179$ & $2543-6125$ & (5)  \\
SDSS J184037.78+642312.3    & $9390\pm140$ & $6.49\pm0.06$ & $0.183$ & $9120\pm140$  & $6.34\pm0.05$ & $0.177$ & $2094-4890$ & (1)  \\
SDSS J111215.82+111745.0    & $9590\pm140$ & $6.36\pm0.06$ & $0.179$ & $9240\pm140$  & $6.17\pm0.06$ & $0.169$ &  $108-2855$ & (3) \\
SDSS J151826.68$+$065813.2  & $9900\pm140$ & $6.80\pm0.05$ & $0.220$ & $9650\pm140$  & $6.68\pm0.05$ & $0.197$ & $1335-3848$ & (3)  \\ 
\noalign{\smallskip}
\hline
\end{tabular}

{\footnotesize References: 
(1) \citet{2012ApJ...750L..28H}; 
(2) \citet{2013MNRAS.436.3573H};
(3) \citet{2013ApJ...765..102H}; 
(4) \citet{2015MNRAS.446L..26K};
(5) \citet{2015ASPC..493..217B}
}
\label{tabla-ELMVs}
\end{table*}  

\citet{2013A&A...557A..19A} derived realistic  configurations   for
low-mass He-core WDs by mimicking the binary evolution  of progenitor
stars. Full details about this procedure are given  in
\citet{2013A&A...557A..19A} and \citet{2014A&A...569A.106C}. 
Binary evolution was assumed to be  fully
nonconservative, and the loss of angular momentum due to  mass loss,
gravitational wave radiation, and magnetic braking was
considered. All of the He-core WD initial models were derived  from
evolutionary calculations for binary systems consisting of an
evolving Main Sequence low-mass component of initially $1 M_{\sun}$
and a $1.4 M_{\sun}$ neutron star as the other component.  A total of
14 initial He-core WD models with stellar masses between  $0.1554$ and
$0.4352 M_{\sun}$ were computed for initial orbital periods at the
beginning of the Roche lobe phase in the range $0.9$ to $300$ d.  In
Table \ref{table1}, we  provide some   relevant characteristics  of
the whole  set of  He-core  WD models.    The   evolution of these
models was computed down to the range of  luminosities of cool WDs,
including the stages of multiple   thermonuclear CNO flashes during
the beginning of cooling branch.  Column 1 of Table \ref{table1} shows
the resulting final stellar masses ($M_{\star}/M_{\sun}$). The second
column corresponds to the total  amount of H contained in the envelope
($M_{\rm H}/M_{\star}$)  at $T_{\rm eff} \approx  10\, 000$ K (at the
final  cooling branch),  and column 3 displays the time spent by the
WD models to cool from $T_{\rm eff} \approx 10\,  000$ K to  $\approx
8000$  K.  Finally,  column 4  indicates the occurrence (or not) of
CNO flashes on the early WD cooling branch.   There exists  a
threshold  in the  stellar mass value  (at $\sim 0.18 M_{\sun}$),
below which CNO flashes on the early  WD cooling branch are not
expected to occur, in agreement with  previous studies
\citep{2000MNRAS.316...84S, 2001MNRAS.323..471A,2004ApJ...616.1124N}. 
Sequences with
$M_{\star} \lesssim 0.18 M_{\sun}$ have thicker H envelopes  and much
longer cooling timescales than sequences with stellar
masses above that mass threshold. To put this in numbers, the H  content and
$\tau$ (the time to cool from $T_{\rm eff} \approx  10\, 000$ K  to
$T_{\rm eff} \approx  8000$ K) for the sequence with  $M_{\star}=
0.1762 M_{\sun}$ are $\sim 4$ and $\sim 22$ times larger,
respectively, than for the sequence with $M_{\star}= 0.1806 M_{\sun}$
(see Table \ref{table1}). Note that in this example,  we are comparing
the properties of two sequences with virtually the same stellar mass
($\Delta M_{\star} \approx 4 \times 10^{-3} M_{\sun}$). The slow
evolution of  the non-flashing sequences is due to that the residual H
burning is the main source of surface luminosity, even  at very
advanced stages of  evolution.  We show in Fig. \ref{figure_02} the
complete set of evolutionary  tracks (final cooling branches) of
our low-mass He-core WDs, along with the seven ELMVs  discovered so
far. We include the location of ELMV stars with $T_{\rm eff}$ and
$\log g$ values derived from 1D model atmospheres (large red circles),
as well as for the case  in which these parameters are corrected for
3D effects (small black circles)  following
\citet{2015arXiv150701927T}.  These parameters are shown in Table
\ref{tabla-ELMVs}. The corrected  effective temperatures and gravities
are extracted directly from  \citet{2015arXiv150701927T}, except for
the star SDSS J1618$+$3854,  for which we use the fitting functions
given by those authors.  Visibly, 3D corrections lower the estimated
1D $T_{\rm eff}$ and $\log g$,  implying lower masses (compare columns
4 and 7 of the Table). 

\section{Stability analysis}
\label{stability}

We analyze the stability pulsation properties of about 7000 stellar
models of He-core, low-mass WDs corresponding to a total of 42
evolutionary  sequences that include three different prescriptions for
the  MLT theory of convection  \citep[ML1, ML2, ML3;
  see][]{1990ApJS...72..335T},  and covering a range of effective
temperatures of   $13\,000\ {\rm K} \lesssim  T_{\rm eff} \lesssim
6\,000$ K and a range of stellar masses of $0.1554 \lesssim
M_{\star}/M_{\sun} \lesssim 0.4352$.  For each model, we assessed the
pulsation stability of radial ($\ell= 0$) and nonradial ($\ell= 1, 2$)
$p$ and $g$ modes with periods  from a range $10\ {\rm s} \leq \Pi
\leq 18\,000$ s for the sequence with $M_{\star}= 0.1554 M_{\sun}$, up
to a range of periods of $0.3\ {\rm s} \leq \Pi \leq 5\,000$ s  for
the sequence of with $M_{\star}= 0.4352 M_{\sun}$.  Certainly, these
ranges of periods  are extremely  wide when compared to the range of
periods observed in ELMVs so far ($100\ {\rm s} \leq \Pi \leq 7\,000$
s).  The reason for considering such wide ranges of periods in our
computations  is to clearly define the theoretical domain of
instability, that is,  to find the long- and short-period  edges of
the instability domains for all the stellar masses and effective
temperatures.

\begin{figure} 
\begin{center}
\includegraphics[clip,width=9 cm]{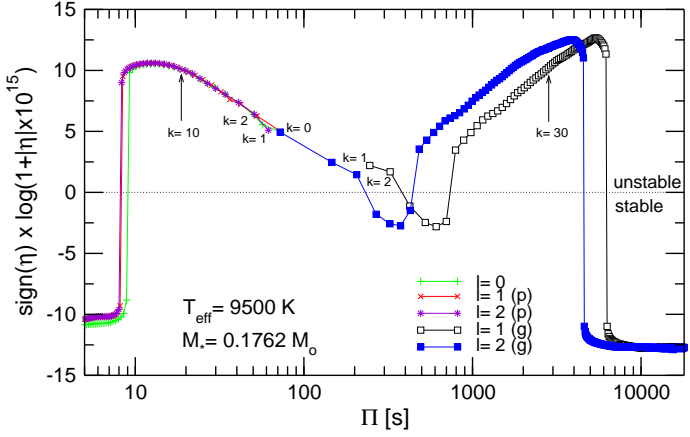} 
\caption{Normalized growth rates $\eta$  (symbols connected with
  continuous lines) for radial ($\ell= 0$) and  nonradial $\ell= 1, 2$
  $p$ and $g$ modes in terms of the pulsation  periods for the
  $0.1762M_{\sun}$ ELM WD template model at  $T_{\rm eff}= 9500$
  K. The large numerical range spanned by $\eta$  is appropriately
  scaled for a better graphical representation. Specific modes (mode
  $p$ with $\ell= 1, k= 10$ and  mode $g$ with $\ell= 1, k= 30$),
  which are analyzed in Fig. \ref{figure_04}, are marked with
  arrows.}
\label{figure_03} 
\end{center}
\end{figure}

We start by discussing the stability properties of a template $0.1762
M_{\sun}$ low-mass He-core WD model with $T_{\rm eff}= 9500$ K  and
ML2 ($\alpha=1.0$). Its location in the $T_{\rm eff} - \log g$ diagram
is displayed in Fig. \ref{figure_02} as a hollow square. These properties
are qualitatively the same for   all the models of our complete set of
evolutionary sequences.  The normalized growth  rate $\eta$ ($\equiv
-\Im(\sigma)/ \Re(\sigma)$,  where $\Re(\sigma)$ and $\Im(\sigma)$ are
the real and the  imaginary part, respectively, of the complex
eigenfrequency  $\sigma$) in terms of pulsation periods $\Pi$ for
overstable $\ell= 0, 1$ and  2 modes corresponding to the selected
model is shown in Fig. \ref{figure_03}.  $\eta > 0$ ($\eta < 0$)
implies unstable (stable) modes. The range of  periods of unstable $p$
modes is not dependent on the harmonic degree, unlike to what happens
in the case of $g$ modes,  for which the period interval of unstable
modes for $\ell= 2$ is shifted  to shorter periods  when compared with
the range of $\ell= 1$ unstable mode periods.  For  radial modes and
$p$ modes, the growth rate reaches  a maximum  value ($\eta_{\rm max}
\sim 4 \times 10^{-5}$) in the  vicinity  of the short-period edge  of
the instability domain. In  other words,  within a given  band of
unstable modes, the excitation  is  markedly  stronger  for modes
characterized  by  short periods (high frequencies).  The opposite
holds for $g$ modes, for which the largest excitation ($\eta_{\rm max}
\sim 4.6 \times 10^{-3}$)  corresponds to the long-period boundary of
the domain of unstable modes.  Radial modes and $p$ modes with
increasing periods (decreasing radial order $k$) are all unstable,
even the lowest order modes, although with
the minimum excitation value ($\eta_{\rm min} \sim
10^{-10}$). Something quite different occurs in the case  of $g$
modes. Specifically, the value of $\eta$ for $g$ modes gradually
decreases for decreasing periods,  until it reaches negative values  for
modes with $k= 3, 4, 5, 6$ which are pulsationally stable.  However,
modes with $k= 1$ and $2$ are again  unstable, although with  very
small growth rates ($\eta \sim 10^{-13}$). In the case of $\ell= 2$,
even the $f$ mode ($k= 0$)  is unstable, as it is clearly documented
in Fig. \ref{figure_03}.  As we will discuss below, stable  nuclear
burning of H plays a role in destabilizing these  low-order $g$ modes
($k= 1, 2$), aside from the strong driving associated to the partial
ionization of H.

\begin{figure} 
\begin{center}
\includegraphics[clip,width=8 cm]{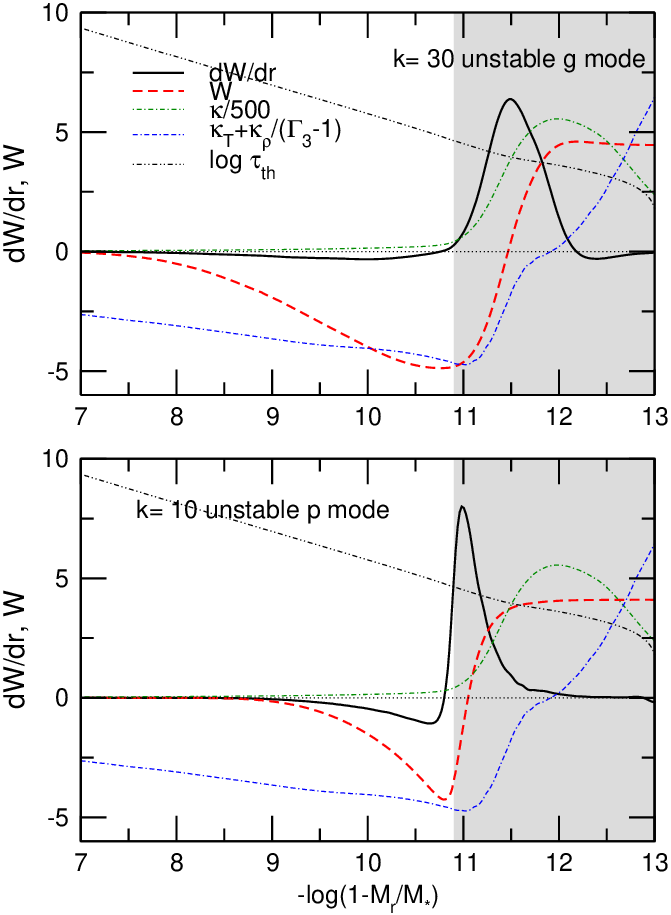}
\caption{The differential work ($dW/dr$) and the running work integral
  ($W$) for the  $g$ mode with $k= 30$ (upper panel)  and the $p$ mode
  with $k= 10$ (lower panel), along with the Rosseland opacity profile
  ($\kappa$),  the  opacity  derivatives,  and the  thermal  timescale
  ($\tau_{\rm  th}$) of  our  $0.1762M_{\sun}$ ELM  WD template  model
  ($T_{\rm eff}= 9500$ K).  The  gray area shows the location of the
  outer convection zone.}
\label{figure_04} 
\end{center}
\end{figure}

We select two  representative unstable  pulsation modes  of the
template   model  in  order   to  investigate   the  details   of  the
driving/damping process. Specifically, we choose two overstable dipole
$(\ell= 1)$ modes, one  of them being a $g$ mode with  $k= 30$ and the
other  one   a  $p$  mode  with   $k=  10$  (marked   with  arrows  in
Fig.    \ref{figure_03}).   In   the    upper   (lower)    panel   of
Fig.  \ref{figure_04}  we display  the  differential work  function
$dW/dr$  and   the  running   work  integral  $W$   \citep[see][for  a
  definition]{1993ApJ...418..855L}  for  the  unstable $g$  mode  ($p$
mode),  characterized by  $\Pi= 2817$  s, $\eta=  1.6  \times 10^{-5}$
($\Pi= 19.07$ s, $\eta= 1.1 \times 10^{-5}$). Note that the scales for
$dW/dr$ and  $W$ are arbitrary.  Also shown are the  Rosseland opacity
($\kappa$)        and       its        derivatives       ($\kappa_{\rm
  T}+\kappa_{\rho}/(\Gamma_3-1)$),  and the  logarithm of  the thermal
timescale ($\tau_{\rm  th}$).  We restrict the figure  to the envelope
region of  the model, where the  main driving and  damping occurs. The
region  that destabilizes  the modes  (where $dW/dr  > 0$)  is clearly
associated with the bump in the  opacity due to the ionization of H at
the outer  convection zone  (gray area in  the Figure),  centered at
$-\log q  \sim 12$ $[q \equiv (1-M_r/M_{\star})]$,  although the
maximum driving comes  from a  slightly more internal  regions ($-\log
q \sim 11.5$ for the  $g$ mode and $-\log  q \sim 11.0$ for the  $p$
mode). The thermal timescale reaches values of  the order of
$10^3-10^4$ s at the driving  region, compatible  with the  longest
excited  period  of the template model, at  $\sim 6200$ s. In the
driving region the quantity $\kappa_{\rm
  T}+\kappa_{\rho}/(\Gamma_3-1)$  is increasing outward, in agreement
with the well  known necessary condition for mode excitation
\citep{1989nos..book.....U}.  For the $g$  mode, the  contributions to
driving at $-\log q$ from $\sim  11$ to $\sim 12$ largely overcome the
weak damping  effects at  $-\log q \lesssim  11$ and $-\log  q \gtrsim
12$, as reflected by the fact that  $W > 0$ at the surface, and so the
mode is globally excited. Similarly, the strong driving experienced by
the $p$ mode  (denoted by positive values of  $dW/dr$ for $11 \lesssim
-\log q \lesssim 11.5$) makes this mode globally unstable.

\subsection{The destabilizing role of H burning}

\begin{table*}
\centering
\caption{The stellar mass, the  mass of H, the evolutionary timescale,
  the  radial order and  harmonic degree,  the $T_{\rm  eff}$-range of
  instability, the average period, the maximum $e$-folding time, and 
  the ratio of the evolutionary timescale to the maximum $e$-folding time, 
  corresponding to unstable  short-period  $\ell=  1,   2$  $g$  
modes  for  which  the
  $\varepsilon$ mechanism strongly  contributes to their destabilization,
  corresponding to  model sequences with $M_{\star} \leq 0.2389 M_{\sun}$ 
computed using  the ML2 prescription of the MLT theory of convection.}
\begin{tabular}{cccccccr}
\hline
\hline
\noalign{\smallskip}
 $M_{\star}/M_{\sun}$  & $M_{\rm H}/M_{\star}$ $[10^{-3}]$ &  $\tau$ $[10^9 {\rm yr}]$   &  
$k\ (\ell)$ & $T_{\rm eff}$ $[$K$]$  & $\langle \Pi\rangle$  $[$s$]$ &   
$\tau_e^{\rm max}$  $[10^9 {\rm yr}]$ & $\tau/\tau_e^{\rm max}$\\
\noalign{\smallskip}
\hline
\noalign{\smallskip}
0.1554       & 25.4  & 3.13            &  2 (1)  & $\lesssim 8500$ & $350$ & $0.07$ & 44.7 \\
             &       &                 &  3 (1)  & $9000-8300$     & $470$ & $0.97$ &  3.2 \\ 
             &       &                 &  2 (2)  & $\lesssim 8100$ & $227$ & $0.12$ & 26.1 \\  
             &       &                 &  3 (2)  & $8600-8360$     & $291$ & $0.20$ & 15.7 \\  
             &       &                 &  4 (2)  & $9000-8800$     & $355$ & $0.33$ & 9.5 \\
\hline
\noalign{\smallskip}
0.1612       & 20.6  & 4.44            &  2 (1)  & $\lesssim 8950$ & $343$ & $0.80$ & 5.6 \\
             &       &                 &  3 (1)  & $\lesssim 9260$ & $448$ & $0.44$ & 10.1 \\
             &       &                 &  1 (2)  & $\lesssim 7150$ & $148$ & $0.03$ & 148.0 \\ 
             &       &                 &  2 (2)  & $\lesssim 8600$ & $220$ & $0.80$ & 5.6 \\ 
             &       &                 &  3 (2)  & $\lesssim 9250$ & $283$ & $0.60$ & 7.4 \\ 
\hline
\noalign{\smallskip}
0.1650       & 18.7  & 5.53            &  1 (1) & $\lesssim 8200$ & $250$ & $0.07$  & 79.0 \\
             &       &                 &  2 (1) & $\lesssim 9500$ & $340$ & $0.17$  & 32.5 \\ 
             &       &                 &  3 (1) & $9500-9020$     & $450$ & $1.30$  & 4.25 \\ 
             &       &                 &  4 (1) & $\lesssim 7800$ & $580$ & $0.80$  & 6.9 \\ 
             &       &                 &  2 (2) & $\lesssim 8950$ & $214$ & $0.50$  & 11.1 \\ 
             &       &                 &  3 (2) & $9400-9300$     & $277$ & $0.50$  & 11.1 \\ 
\hline
\noalign{\smallskip}
0.1706       & 16.3  & 6.59            &  1 (1) &  $9660-9600$    & $255$ & $0.05$ & 131.8\\
             &       &                 &  2 (1) & $\lesssim 9979$ & $350$ & $1.05$ & 6.3 \\
             &       &                 &  3 (1) & $\lesssim 7750$ & $480$ & $2.50$ & 2.6 \\
             &       &                 &  1 (2) & $\lesssim 9060$ & $151$ & $0.01$ & 659 \\
             &       &                 &  2 (2) & $\lesssim 9650$ & $223$ & $1.95$ & 3.4 \\
             &       &                 &  3 (2) & $7730-7180$     & $310$ & $0.58$ & 11.4 \\   
\hline
\noalign{\smallskip}
0.1762       & 14.5 & 7.56             &  1 (1) & $\lesssim 9100$ & $247$ &   $1.40$  & 5.4 \\
             &      &                  &  2 (1) & $\lesssim 10\,000$& $320$ &  $0.20$ & 37.8 \\
             &      &                  &  3 (1) & $\lesssim 8700$ & $470$ &   $0.70$   & 10.8 \\
             &      &                  &  4 (1) & $8900-8700$       & $550$ &   $0.09$ & 84 \\
             &      &                  &  5 (1) & $9200-9150$       & $620$ &   $0.06$ & 126 \\
             &      &                  &  1 (2) & $\lesssim 8300$   & $140$ &   $0.02$ & 378 \\
             &      &                  &  2 (2) & $\lesssim 9900$   & $220$ &   $0.40$ & 18.9 \\
             &      &                  &  3 (2) & $8300-7700$       & $297$ &   $0.25$ & 30.4 \\
\hline
\hline
\noalign{\smallskip}
0.1806       & 3.68 & 0.34             &  1 (1) & $\lesssim 10\,500$   & $270$ &  $0.05$ & 6.8 \\
             &      &                  &  2 (1) & $10\,200-9700$       & $355$ &  $0.02$ & 17 \\  
             &      &                  &  1 (2) & $\lesssim 10\,500$   & $178$ &  $0.40$ & 0.85\\  
\hline
\noalign{\smallskip}
0.1863       & 4.36  & 0.37            &  1 (1) & $\lesssim 10\,500$ &  $286$  &  $0.50$ & 0.74 \\
             &       &                 &  2 (1) & $10\,200-9000$     &  $355$  &  $0.60$ & 0.62\\
             &       &                 &  1 (2) & $\lesssim 10\,500$ & $168$   &  $0.57$ & 0.65 \\
             &       &                 &  2 (2) & $10\,200-9000$     & $219$   &  $0.68$ & 0.54\\
\hline
\noalign{\smallskip}
0.1917       & 4.49  & 0.35            &  1 (1) & $\lesssim 10\,500$ &  $280$  & $0.45$ & 0.78\\
             &       &                 &  2 (1) & $\lesssim 10\,220$ &  $340$  & $0.72$ & 0.49\\
             &       &                 &  1 (2) & $\lesssim 10\,500$ &  $160$  & $0.57$ & 0.61\\
             &       &                 &  2 (2) & $\lesssim 8900$    &  $210$  & $2.60$ & 0.13\\
\hline
\noalign{\smallskip}
0.2019       & 3.80  & 0.32            &  1 (1) & $\lesssim 10\,600$ & $263$  & $0.09$ & 3.6\\
             &       &                 &  1 (2) & $\lesssim 10\,590$ & $155$  & $0.25$ & 1.28\\
\hline
\noalign{\smallskip}
0.2389       & 3.61  & 0.62            &  1 (1) & $\lesssim 10\,700$ & $200$  & $0.04$  & 15.5\\
             &       &                 &  2 (1) & $10\,500-7620$        & $300$  & $0.15$  & 4.1\\
             &       &                 &  1 (2) & $\lesssim 10\,700$ & $120$  & $0.06$  & 10.3\\
             &       &                 &  2 (2) & $10\,500-9000$        & $180$  & $0.07$  & 8.9\\
\noalign{\smallskip}
\hline
\hline
\end{tabular}
\label{table-epsilon}
\end{table*}

\begin{figure} 
\begin{center}
\includegraphics[clip,width=9 cm]{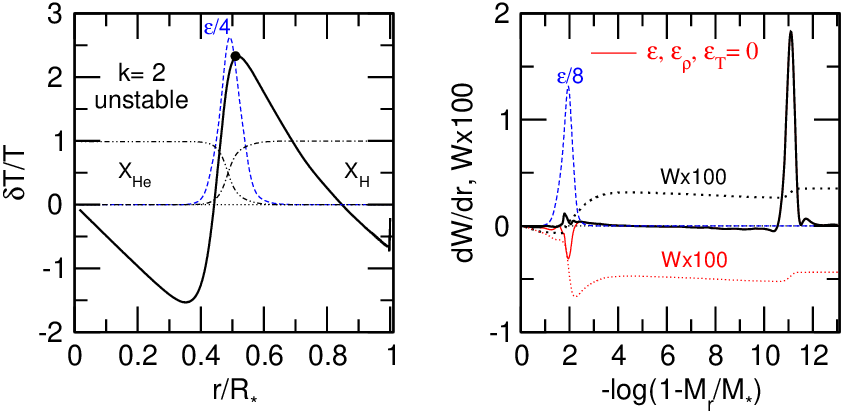} 
\caption{Left panel depicts the Lagrangian perturbation of temperature
  ($\delta  T  /T$) along  with  the  scaled  nuclear generation  rate
  ($\epsilon$) and the H and He chemical abundances ($X_{\rm H},X_{\rm
    He}$) in terms of the  normalized stellar radius, for the unstable
  $\ell=  1,  k= 2$  $g$  mode ($\Pi=  325$  s)  corresponding to  our
  template model.  The  black dot marks the location  of the outermost
  maximum of  $\delta T  /T$. The right  panel shows  the corresponding
  differential work  function ($dW/dr$) in terms of  the mass fraction
  coordinate  for the  case in  which the  $\varepsilon$  mechanism is
  allowed to  operate (solid  black curve) and  when it  is suppressed
  (solid red curve). Also shown are the scaled running work integrals,
  $W$ (dotted curves).}
\label{figure_05} 
\end{center}
\end{figure}

\citet{2014ApJ...793L..17C}  have  explored the  impact  of
stable H burning  on the pulsational stability properties  of the same
models of low-mass He-core WDs analyzed here. They found that, besides
a dense  spectrum of unstable radial  modes and nonradial  $g$ and $p$
modes  driven by  the  $\kappa-\gamma$ mechanism  due  to the  partial
ionization of H in the  stellar envelope, some unstable $g$ modes with
short pulsation  periods are  also destabilized by  H burning  via the
$\varepsilon$    mechanism    \citep{1989nos..book.....U}   of    mode
driving. \citet{2014ApJ...793L..17C} speculate  that the short periods
at $\Pi  \sim 108$ s and  $\Pi \sim 134$  s detected in the  ELMV star
SDSS   J1112$+$1117   \citep{2013ApJ...765..102H}  could   be
excited by  this mechanism. If  true, this could constitute  the first
evidence  of the existence  of stable  H burning  in cool WD  stars.  These
interesting  results rely,  however,  in the  reality  of those  short
pulsation periods, something that  needs additional observations to be
confirmed (J. J. Hermes, private communication).

\begin{figure*} 
\begin{center}
\includegraphics[clip,width=17 cm]{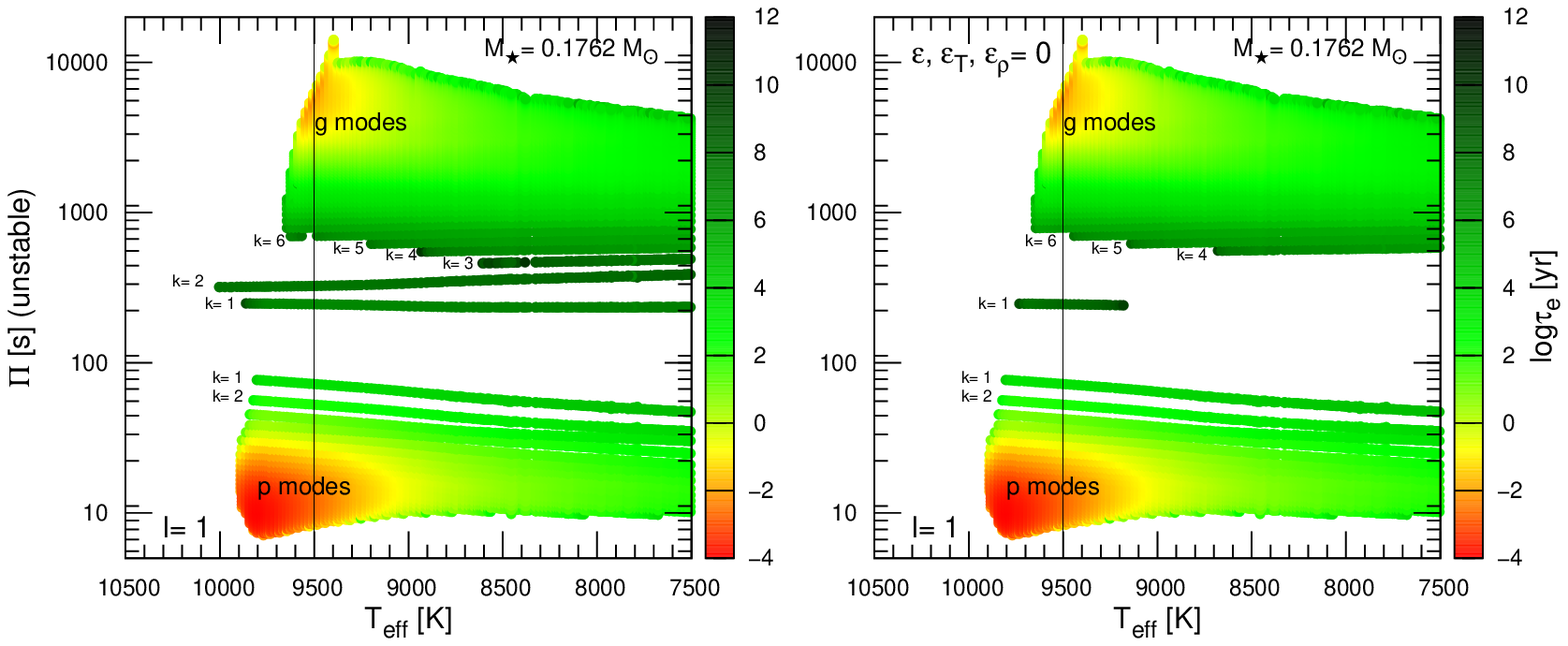} 
\caption{Left panel: unstable $\ell=  1$ mode periods ($\Pi$) in terms
  of  the effective  temperature, corresponding  to the  ELM  WD model
  sequence with  $M_{\star}= 0.1762 M_{\sun}$.  Color coding indicates
  the value of the logarithm  of the $e$-folding time ($\tau_{\rm e}$)
  of  each  unstable mode  (right  scale).  The  vertical  line
  indicates the  effective temperature of the  template model analyzed
  in Fig. \ref{figure_03}.  Right panel: same as left  panel, but for
  the  case in  which the  action  of the  $\varepsilon$ mechanism  is
  neglected in the stability computations.}
\label{figure_06} 
\end{center}
\end{figure*}

In order to  show how the $\varepsilon$ mechanism  acts to destabilize
low-order $g$-modes in our models,  we restrict the description to the
$k= 2$  mode of  our selected template  model. This mode  is unstable,
with  $\eta= 4.76 \times  10^{-14}$. Admittedly,  this growth  rate is
extremely small,  and one  can wonder if  this mode,  being marginally
unstable,  is  able  to  reach   amplitudes  large  enough  as  to  be
observable. To answer that question we have to examine the $e$-folding
time,  defined  as $\tau_{\rm  e}=  1/|\Im(\sigma)|$.  For this  mode,
$\tau_{\rm e}= 3.44 \times 10^{7}$ yr. This is an estimate of the time
it  would take  the mode  to reach  amplitudes large  enough as  to be
observable.  It has  to be  compared with  the  evolutionary timescale
($\tau$) which represents  the time that the model  spends evolving in
the regime  of interest. In this  case, we have $\tau  \sim 7.6 \times
10^9$ yr\footnote{More  precisely, this is the time  that the template
  model  takes to  cool from  $T_{\rm eff}  \sim 10\,000$  K  to $\sim
  8\,000$ K.}.  Since $\tau_{\rm  e} \ll \tau$,  we conclude  that the
mode has plenty of time to  develop large amplitudes while the star is
slowly cooling at that effective temperature regime.

In order to assess the role  that stable H burning has in the
destabilization of the $k= 2$ $g$  mode of the template model, we have
redone the  stability computations, but  this time by  suppressing the
action of this destabilizing agent. Specifically, we force the nuclear
energy production rate,  $\epsilon$, and their logarithmic derivatives
$\epsilon_{\rm T}= (\partial \ln \epsilon/\partial \ln T )_{\rho}$ and
$\epsilon_{\rm \rho}= (\partial  \ln \epsilon/\partial \ln \rho )_{\rm
  T}$ to be zero in the  pulsation equations. The results are shown in
the  right  panel  of  Fig. \ref{figure_05},  that  shows  the
differential  work function ($dW/dr$)  in terms  of the  mass fraction
coordinate  for  the case  in  which  the  $\varepsilon$ mechanism  is
allowed  to operate  (solid black  curve)  and when  it is  suppressed
(solid red curve).  Also shown are the scaled  running work integrals,
$W$  (dotted curves).   The  left  panel of  the  Figure displays  the
Lagrangian perturbation of temperature, $\delta T /T$. The peak of the
(scaled) nuclear generation rate $\epsilon$ at $r/R_{\star} \sim 0.45$
marks  the  location of  the  H-burning  shell  at the  He/H  chemical
interface.  We emphasize  the  position of  the  outermost maximum  of
$\delta T /T$ with a black dot. The $\varepsilon$ mechanism behaves as
an efficient filter of modes that provides substantial driving only to
those $g$  modes that have their  largest maximum of $\delta  T /T$ in
the      narrow       region      of      the       burning      shell
\citep{1986ApJ...306L..41K}.  The  $g$  mode  with  $k=  2$  met  this
condition.

In  the  case in  which  the  $\varepsilon$  mechanism is  taken  into
account, there is  appreciable driving ($dW/dr > 0$)  at the region of
the H  burning shell ($-\log q  \sim 2$), as can be appreciated from
the right panel of Fig. \ref{figure_05}. The  destabilizing
effect of nuclear  burning  adds up  to  the excitation  due  to  the
H  partial ionization  at the  envelope  ($-\log q  \sim  11$), and
the mode  is globally unstable.  Both contributions  of driving are
also visible as positive slopes of the work  integral at those
locations of the model.  When  we suppress  the  $\varepsilon$
mechanism  (red curves),  strong damping takes place at the region  of
the burning shell. In this case, the driving  due to the  H partial
ionization  at the envelope  is not able  to overcome the  damping at
that region,  and the  mode results pulsationally stable.   We can
conclude that for  this specific model, the  $g$  mode  with  $k=  2$
is  globally  unstable  thanks  to  the destabilizing effect of the
H-burning shell through the $\varepsilon$ mechanism.

A  more  general  and  comprehensive   perspective  of  the  role  of
the $\varepsilon$  mechanism in our  pulsation models  can be
achieved by examining the properties  of our template model for  the
full range of effective     temperatures.      In     the     left
panel     of Fig. \ref{figure_06}  we show the instability
domain of $\ell= 1$ periods in terms of the  effective temperature for
the ELM WD model sequence  with $M_{\star}=  0.1762  M_{\sun}$. The
palette of  colors (right scale) indicates the value  of the logarithm
of the $e$-folding time (in years) of each unstable  mode.  As can be
seen, many unstable pulsation modes exist, which are  clearly grouped
in the two families, one  of them  corresponding to  long periods  and
associated  with $g$ modes, and the other one  characterized by short
periods and belonging to  $p$   modes.  Most  of   these  modes  are
destabilized   by  the $\kappa-\gamma$ mechanism  acting at the
surface H partial ionization zone.  The strongest  excitation, that
is,  the smallest  $e$-folding time (light red and yellow zones), is
found for high-order $g$ and $p$ modes,  with periods  in the  ranges
$3000-10\,000$ s  and $7-30$  s, respectively, and effective
temperatures  near the hot boundary of the instability islands
($T_{\rm eff} \sim 9600-9800$ K). Similar results, although  with
shorter unstable  $g$-mode periods  ($2000-10\,000$) s, are  obtained
for   $\ell=  2$  (not  shown).  The   right  panel  of Fig.
\ref{figure_06}  shows   the  results  of  our  stability
computations     when    we     shut     down    the     $\varepsilon$
mechanism.  Interestingly enough, the  $k =  2$ and  $k= 3$  $g$ modes
become  stable  for  the  complete  range  of  effective  temperatures
analyzed,  and they  do not  appear  in this  plot. Something  similar
happens  with the modes  with $k=1$,  $k =  4$ and  $k= 6$  in certain
ranges of $T_{\rm eff}$. We can conclude that these modes are excited
to a great extent by the $\varepsilon$ mechanism through the H-burning
shell.  We note  that  high-order  $g$ modes  are  insensitive to  the
effects  of nuclear  burning,  and  the same  holds  for the  complete
spectrum of $p$ modes and radial  modes. 

In Table \ref{table-epsilon} we present the short-period  $\ell=
1,   2$  $g$   modes  for  which  the $\varepsilon$ mechanism strongly
contributes to  their destabilization, corresponding to sequences with stellar 
masses below $0.2389 M_{\sun}$.  Note that the number of 
$\varepsilon$-destabilized
modes is larger for  sequences with masses $\leq 0.1762 M_{\sun}$, by
virtue that these models have thick H envelopes and, as a result, they
are able to sustain an  intense H nuclear burning. For models with
$M_{\star} \geq 0.1806 M_{\sun}$, nuclear burning is much weaker, but
still able to contribute  to the driving of $g$ modes with radial
order $k= 1, 2$. Note, however, that in the case of models with
$M_{\star}= 0.1863 M_{\sun}$ and  $M_{\star}= 0.1917 M_{\sun}$, the
$e$-folding times are shorter than the evolutionary timescale, with
ratios $\tau/\tau_e^{\rm max} < 1$.  These modes, although formally
unstable, do not have enough time   as to reach observable amplitudes
while the star is crossing  the $T_{\rm eff}$ interval $10\,000-8000$
K. For masses above $0.2389 M_{\sun}$ (not shown) only modes with 
$\ell= 1, 2$ and $k= 1$ are $\varepsilon$-destabilized modes.

\subsection{Characterizing the blue edge of 
the theoretical ELMV instability strip}
\label{strip}

\begin{figure} 
\begin{center}
\includegraphics[clip,width=9 cm]{fig_07.eps} 
\caption{The $T_{\rm eff}-\log  g$ diagram showing the blue  edge of
  the ELMV instability domain  for the cases of radial  modes (black
  solid line),  nonradial dipole  $g$  modes (solid  red line),
  nonradial quadrupole $g$  modes (dashed red  lines), and nonradial
  dipole and quadrupole  $p$ modes  (blue solid  line),  for the  case
  in  which stellar models are computed using the ML2 version of the
  MLT theory.  The  known ELMVs are also depicted. Filled red circles
  correspond to the location of the stars using 1D model 
  atmospheres, filled black circles mark their location according to 3D 
  model atmospheres, and hollow blue circles are 
  associated to low-mass WDs  observed not to vary.}
\label{figure_07} 
\end{center}
\end{figure}

\begin{figure*} 
\begin{center}
\includegraphics[clip,width=16 cm]{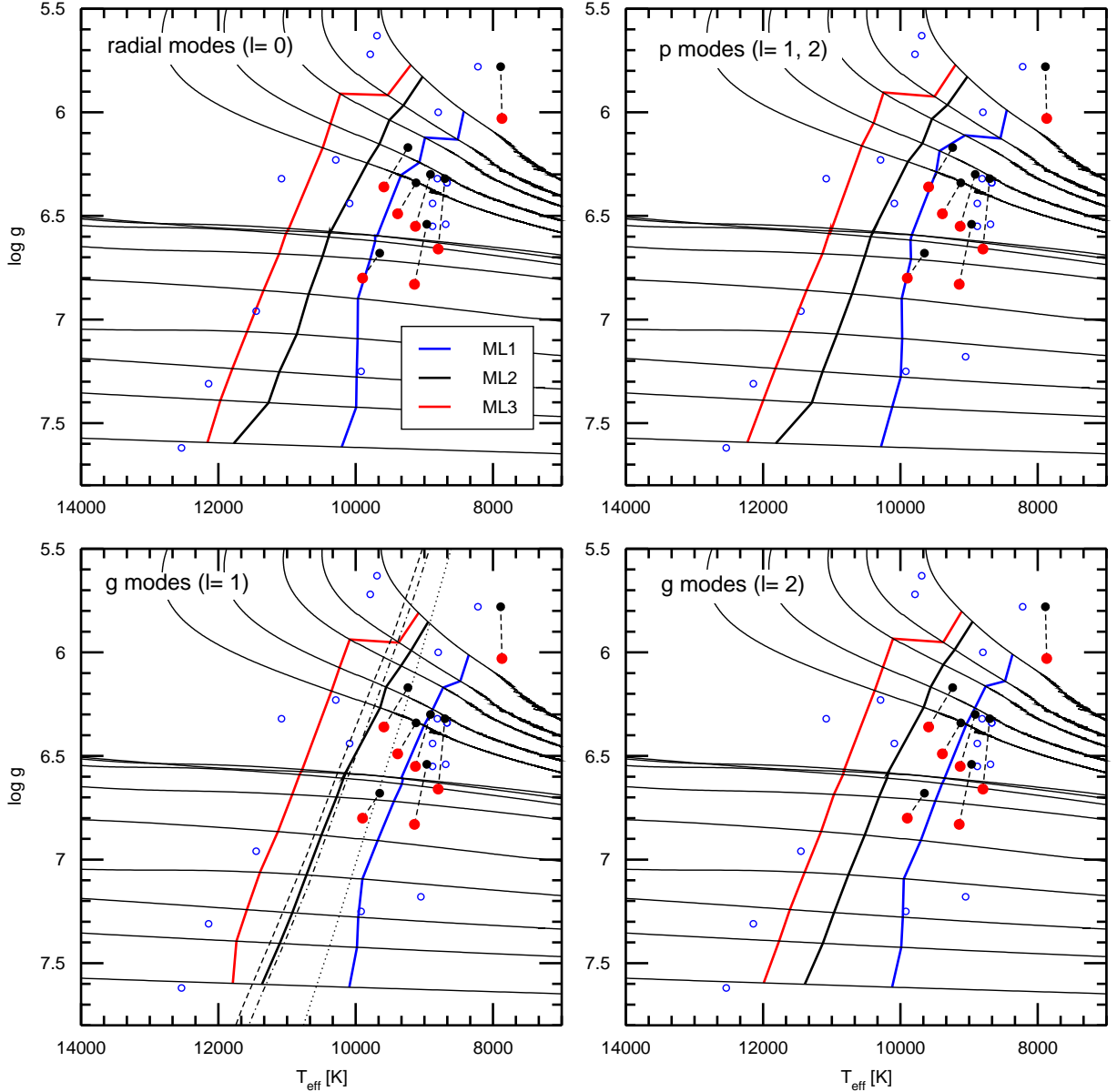} 
\caption{$T_{\rm eff}-\log g$ diagrams displaying our low-mass He-core
  WD  evolutionary tracks along with the blue (hot) edge of the  ELMV
  instability strip for radial ($\ell= 0)$ and nonradial ($\ell= 1,
  2$) $p$ and $g$ modes, corresponding to different versions of the
  MLT theory of convection: ML1 (blue), ML2 (black), and ML3 (red).
  Again,  the  known ELMVs and the stars observed not to vary are 
  also depicted. 
  In the case of $\ell= 1$ $g$ modes we have included the  blue edges
  computed with the TDC treatment (dashed black line) and the FC
  approximation (dot dashed black line), and also the red edge
  (dotted black line) of the instability strip from  
  \citet{2013ApJ...762...57V}.}
\label{figure_08} 
\end{center}
\end{figure*}

Here,  we examine  the  location  of the  instability  domains of  our
low-mass He-core  WDs  for radial and  nonradial $g$  and $p$ modes on
the $T_{\rm eff} - \log  g$ plane. The locus  of the blue (hot) edge
of instability is illustrated in Fig. \ref{figure_07} for the
case in which the surface convection in the equilibrium models is
treated according to the ML2 $(\alpha= 1)$ version of the MLT theory.
A note about the way in which the blue edge is defined in this work is
in order.  Generally, as the models of our  evolutionary sequences
cool,  the first $g$ modes to became  unstable are  those of  low
radial  order  ($k= 1,  2$).  In most cases, the $\varepsilon$
mechanism plays a crucial role  in the destabilization of these modes
(see Figs. \ref{figure_05} and \ref{figure_06}). At
something  lower $T_{\rm  eff}$s,   higher order modes  ($k=  6, 7,
8, \ldots$) are destabilized,  while intermediate order modes ($k= 3,
4, 5$) remain  pulsationally stable at  those effective  temperatures
(Fig.  \ref{figure_06}). This is particularly notorious in our
less massive ($M_{\star} \leq 0.1762 M_{\sun}$) sequences.  So,  in
defining the  blue  edge of instability for  $g$ modes, we adopt  the
effective temperature at which  the bulk of modes ($k \gtrsim 6$)
become unstable\footnote{If, instead, we were adopting the $T_{\rm
    eff}$ at  which modes with  $k= 1, 2$  become unstable, then  the
  blue  edges would  be somewhat  hotter.}. In  the  case of radial
modes and  nonradial $p$ modes, there is  no ambiguity since  these
modes  are destabilized  gradually,  starting from  the lowest radial
orders ($k= 0, 1, 2, 3, \ldots$ for radial modes and  $k= 1, 2, 3,
\ldots$ for $p$ modes).

Fig.  \ref{figure_07} shows  that  the blue  edges associated  to
radial  modes and nonradial  $p$ modes  are $\sim  200$ K  hotter than
those corresponding  to $g$  modes. This means  that radial  modes and
nonradial $p$ modes  are first destabilized as the  models cool, as it
can be clearly appreciated from Fig. \ref{figure_06}.  We found
also that  the blue edge of  radial modes is slightly  cooler than
that of the $p$ modes, and that the blue edge for the $p$ modes is
largely  independent on the harmonic degree. On the other hand,  the
blue edge of $g$ modes is weakly  sensitive to  the $\ell$  value,
being  up to  $\sim  45$ K  hotter for $\ell= 2$ than for $\ell= 1$.

\begin{figure*} 
\begin{center}
\includegraphics[clip,width=17 cm]{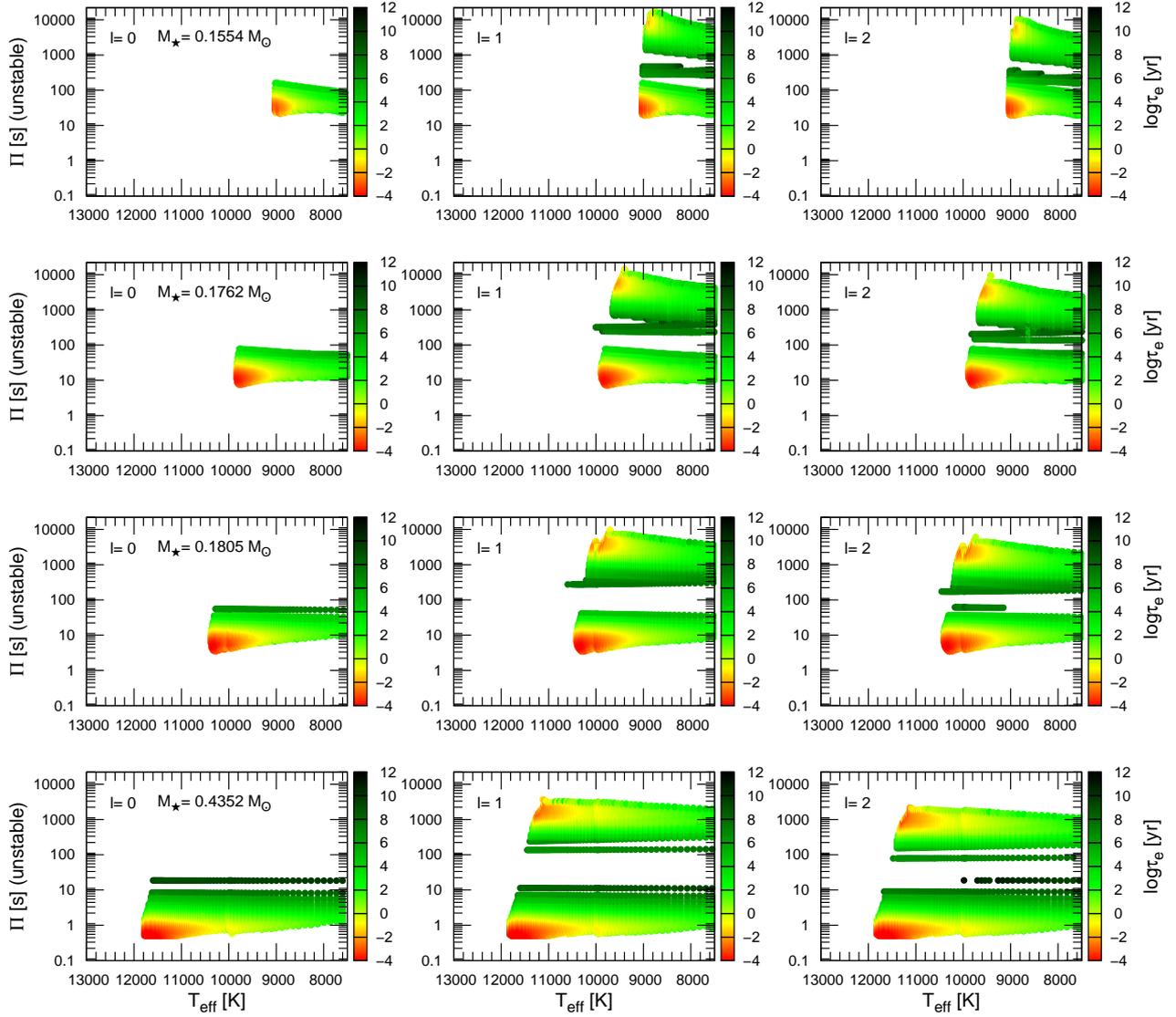} 
\caption{Unstable mode periods ($\Pi$) for $\ell= 0$ (left panels),
  $\ell= 1$ (middle panels), and $\ell= 2$ (right panels) in terms of
  the effective  temperature, corresponding  to the  ELM  WD model
  sequences  constructed using the ML2 version of the MLT theory of
  convection, and  with stellar masses (from top to bottom)
  $M_{\star}= 0.1554 M_{\sun}$,  $M_{\star}= 0.1762 M_{\sun}$,
  $M_{\star}= 0.1805 M_{\sun}$, and  $M_{\star}= 0.4352 M_{\sun}$.
  Color coding indicates the value of the logarithm  of the
  $e$-folding  time ($\tau_{\rm e}$) of  each  unstable mode  (right
  scale).}
\label{figure_09} 
\end{center}
\end{figure*}

The dependence of the blue edges of instability with the convective
efficiency adopted in the equilibrium models is documented in
Fig. \ref{figure_08}, in which  we show the $T_{\rm eff}-\log g$
diagrams displaying our low-mass He-core WD  evolutionary tracks along
with the blue (hot) edge of the  ELMV instability strip for radial
($\ell= 0)$ and nonradial ($\ell= 1, 2$) $p$ and $g$ modes,
corresponding to different versions of the  MLT theory of convection:
ML1 (blue), ML2 (black), and ML3 (red). 
As expected, the blue edge in the case of the ML3 version is hotter
than  for the ML2 version, and it in turn is hotter than the ML1
prescription.  The shift in the effective temperature of the blue
edges for   the different versions of the MLT is between $\sim 600$ K
and $\sim 1300$ K.  The main result exhibited by this figure
is that only the  ML2 and ML3 prescriptions of the MLT actually 
account for
all the observed ELMV stars  (filled red and black circles),
regardless   whether or not 3D model atmosphere corrections are
considered to estimate  $T_{\rm eff}$ and $\log g$. Interestingly,  
no ELMV is found to be hotter than the blue edge associated to the ML2 
version, but it can be due to the small sample of stars.

Our blue edge for $\ell= 1$ $g$ modes with ML2 is in excellent
agreement with the  blue edges derived by \citet{2013ApJ...762...57V},
shown  in the lower left  panel of  Fig. \ref{figure_08} with  a
black dashed  line in the case of a time dependent convection
treatment (TDC),  and with a black dot-dashed line when those authors
use the
FC approximation.  Note that the agreement between our computations
and those of  \citet{2013ApJ...762...57V} breaks down for masses
smaller than the limit mass  $\sim 0.18 M_{\sun}$, below which CNO
flashes on the early  WD cooling branch are not expected to occur. We
have also  included an estimation for the red edge, as proposed by
\citet{2013ApJ...762...57V} (black dotted
line). This estimation is based on the atmosphere energy leakage 
argument elaborated by \citet{1985ApJ...297..544H}. 
It is apparent that the proposed red edge from \citet{2013ApJ...762...57V}
does not describe the observations.

We now explore the ranges of periods of unstable modes and their
dependence with stellar mass, the effective temperature and the
version of the  MLT theory employed. Fig. \ref{figure_09} shows
the unstable radial and  nonradial $p$ and $g$ modes on the $T_{\rm
  eff} - \Pi$ plane for the evolutionary  sequences with
$M_{\star}/M_{\sun}= 0.1554, 0.1762, 0.1805$ and $0.4352$.   The
periods of unstable nonradial modes for each sequence are clearly
grouped in two separated regions, one of them characterized by short
periods and corresponding to $p$ modes, and the  other one
characterized by long periods and associated to $g$ modes. In the case
of radial modes, there is a single instability region with periods
very similar to those of the nonradial $p$ modes. In the case of
$\ell= 2$, there is the $f$ mode  in between the regions of $p$ and
$g$ modes. The longest excited periods for $g$  modes reach values up
to $\sim 15\,500$ s ($\ell= 1$) and $\sim 10\,000$ s ($\ell= 2$) for
the lowest-mass  sequence  ($M_{\star}/M_{\sun}= 0.1554$), and these
numbers drastically decrease to $\sim 3500$ s ($\ell= 1$) and  $\sim
2000$ s ($\ell= 2$) for the most massive sequence
($M_{\star}/M_{\sun}= 0.4352$).  The shortest excited periods, in
turn, range from $\sim 285$ s ($\ell= 1$) and $\sim 185$ s ($\ell= 2$)
for $M_{\star}/M_{\sun}= 0.1554$, to $\sim 135$ s ($\ell= 1$)  and
$\sim 80$ s ($\ell= 2$)  for $M_{\star}/M_{\sun}= 0.4352$.  So, the
longest and shorter excited periods of $g$-modes are  larger for lower
$M_{\star}$ and smaller $\ell$. In the case of radial  modes and
nonradial $p$ modes, the shortest excited periods range from $\sim 19$
s ($M_{\star}/M_{\sun}= 0.1554$) up to $\sim 0.5$ s
($M_{\star}/M_{\sun}= 0.4352$).  Notably, the shortest  excited
periods  of $p$ modes are insensitive to the value of $\ell$.  On the
other hand, the longest excited periods (which also are insensitive to
the value of $\ell$) go from $\sim 160$ s ($M_{\star}/M_{\sun}=
0.1554$)  to $\sim 12$ s ($M_{\star}/M_{\sun}= 0.4352$). We conclude
that the longest and shortest excited periods of $p$ modes and radial
modes are larger for lower $M_{\star}$ and they do not depend on
$\ell$.  We did not find   qualitative differences  in the
characteristics of the instability  domains of radial and nonradial
$p$ modes, except for a very small  shift in the effective temperature
of the blue edges, as mentioned before.

Regarding the strength of the mode instability, we found that
generally the  most unstable modes (that is, with the shortest
$e$-folding times) are those characterized by  high radial orders.  As
for the dependence of the destabilization of modes with $T_{\rm eff}$,
we  found that the most unstable pulsation  modes correspond to
stellar models located near to the blue edge of  instability.  The
modes gradually become less unstable  as the model cools. All these
properties are clearly illustrated  in Fig. \ref{figure_09}.  In
the context of our non-adiabatic calculations, which assume the FC
approximation,  the red edge of the instability domain (that is, the
effective temperature at  which the pulsations stop) is located at
about $T_{\rm eff}= 6000-5000$ K (not shown  in
Fig. \ref{figure_09}), much lower than the $T_{\rm eff}$
of the coolest known ELMV star (SDSS J2228+3623, $T_{\rm eff}\sim 7900$ K). 
This disagreement can not be attributed,
however, to the use the FC approximation since an identical  result is
found by \citet{2013ApJ...762...57V} using a TDC treatment (see their Figure
3)\footnote{This red edge, which emerges from the nonadiabatic computations 
of \citet{2013ApJ...762...57V}, should not be confused with the estimation 
of the red edge carried out by the same authors on the basis of  
the atmosphere energy leakage argument (Fig. \ref{figure_09}).}. 
Similarly, \citet{2012A&A...539A..87V} found $T_{\rm eff} \lesssim 6000$ K
for the red edge of ZZ Ceti stars. Clearly, a missing
physical mechanism is at work in real stars, that quenches the
pulsations at  much higher effective temperatures.

Finally, we have examined the dependence of the longest and shortest
excited  periods with the prescription of the MLT theory employed. For
sequences  with $M_{\star} \leq 0.1762 M_{\sun}$, we found that the
longest excited period of $g$ modes is substantially larger for higher
convective efficiency. In particular, in the case of the sequence
with $M_{\star}/M_{\sun}= 0.1554$, we found that the longest excited
period is $\sim 13\,600$ s for ML1, $\sim 15\,500$ s for ML2,  and
$\sim 17\,600$ s for ML3. On the other hand, for sequences with
$M_{\star} \geq 0.1806 M_{\sun}$ the trend is the opposite: the
longest excited period of $g$ modes is shorter  for higher convective
efficiency, although the differences are small. For instance,  for the
sequence  with $M_{\star}/M_{\sun}= 0.4352$, we found  that the
longest unstable period is $\sim 3350$ s (ML1), $\sim 3300$ s (ML2),
and  $\sim 3150$ s (ML3). Regarding the shortest excited periods  for
$g$ modes, we do not found an appreciable dependence with  the
convective efficiency of the models. In the case of $p$ modes  and
radial modes, we found that the largest and shortest excited periods
are rather insensitive to the version of the MLT employed, having
however  a weak trend of higher shortest and longest unstable periods
with higher convective  efficiency.

\section{Comparison with the observed ELMVs}
\label{obs}

Having shown that our theoretical predictions are in good agreement
with  the position of the ELMVs in the diagram $T_{\rm eff}- \log g$
---provided that the stellar models are computed with the ML2 version
of the  MLT theory of convection---, in this Section want to compare
the theoretical ranges of periods associated to unstable modes  with
the pulsation periods exhibited by the observed stars.  In Table
\ref{tabla-ELMVs} we show the main  spectroscopic data available for
the seven ELMV stars known up to now. We include the values of $T_{\rm
  eff}$ and  $\log g$ derived from 1D model atmospheres and the
stellar mass $M_{\star}$ computed from the tracks of
\citet{2013A&A...557A..19A}, and  also in the case in which $T_{\rm
  eff}$ and  $\log g$ are corrected by 3D effects,  following
\citet{2015arXiv150701927T}. Here, we first shall adopt   the
effective temperatures and gravities of ELMV stars derived from 1D
model atmospheres, and next we will consider the case in which the
values  are corrected by 3D effects.  

It should be noted that, for ELM stars, only $T_{\rm eff}$ and $\log g$ 
can be directly constrained from observations and model atmospheres,
and not their stellar mass. In order to get the mass, it is necessary
to assume some evolutionary stage, because many low-mass WD 
evolutionary tracks overlap at different stages due to the 
developments of CNO flashes \citep[see Fig. 2 of][]{2013A&A...557A..19A}.
Because of this, the masses of ELMVs could be off by 
$0.10 M_{\sun}$ or more if the ELM WD is experiencing a CNO flash. 

\subsection{Using  $T_{\rm eff}$ and $\log g$ derived from 1D model atmospheres} 
\label{1D}

\begin{figure} 
\begin{center}
\includegraphics[clip,width=9 cm]{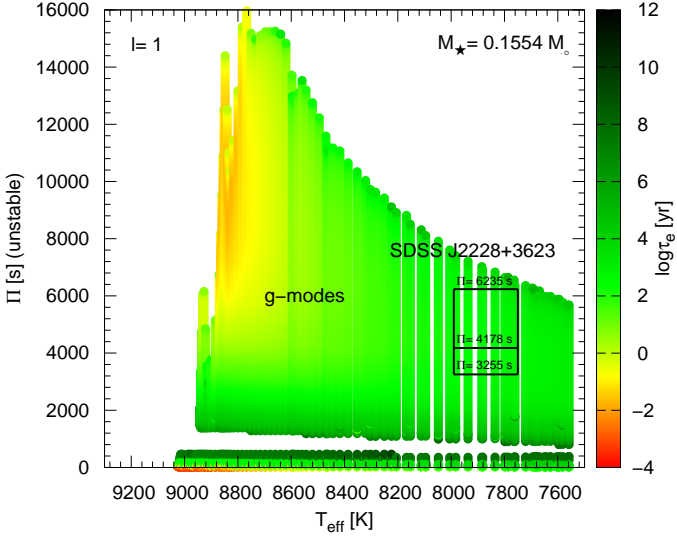} 
\caption{The periods of unstable $\ell= 1$ modes in terms of the
  effective temperature, with the  palette of  colors (right scale)
  indicating the value  of  the logarithm of the $e$-folding time (in
  years), corresponding to the sequence with  $M_{\star}= 0.1554
  M_{\sun}$. Also shown are the pulsation periods of the ELMV star
  SDSS J2228$+$3623 (horizontal segments). The adopted value of
  $T_{\rm eff}$ (and its uncertainties) is that derived from 1D model
  atmospheres  (see Table\ref{tabla-ELMVs}). The gaps for 
  $T_{\rm eff} \lesssim 8200$ K in the unstable models are due to an 
  insufficiently fine grid.}
\label{SDSSJ222859} 
\end{center}
\end{figure}

We start by considering the ELMV star SDSS J2228$+$3623,   the coolest
and less massive object of the class detected to date ($T_{\rm eff}
\sim 7900$ K and  $M_{\star}\sim 0.15 M_{\sun}$). The pulsations of
this star were discovered by \citet{2013MNRAS.436.3573H}. The fact
that this star is so cool as compared with the six warmer pulsating
ELMVs rises the question of whether this star is an    authentic  ELMV
star or is instead a more massive pre-ELM WD  that is looping through
the  $T_{\rm eff} - \log g$ diagram prior to settling on its final WD
cooling track \citep{2013MNRAS.436.3573H}. 
The hypothesis that this star
might be a pre-WD is interesting, even taking into account that the 
evolution of the pre-WDs is much faster than that of the ELM WDs, and 
therefore there are far fewer opportunities of observing it. This issue 
has been examined by \citet{2014A&A...569A.106C},  but without conclusive
results. In Fig. \ref{SDSSJ222859} we show the theoretical unstable
$\ell= 1$ mode periods corresponding to the  evolutionary sequence of
$M_{\star}= 0.1554 M_{\sun}$, the closest stellar mass of our grid  to
the mass inferred for this star. We also include the pulsation periods
of SDSS J2228$+$3623 at 3254.5 s, 4178.3 s, and 6234.9 s. The three
periods are well accounted for by the theoretical computations, being
the longest period ($6234.9$ s) quite close to the theoretical upper
limit of unstable mode  periods at the lower limit of $T_{\rm eff}$ of
this star  ($\Pi_{\rm max} \sim 6800$ s). Note that for the effective
temperatures of interest,  $g$ modes are still quite unstable, with
$e$-folding times of roughly  $10^3-10^4$ yrs. 

\begin{figure} 
\begin{center}
\includegraphics[clip,width=9 cm]{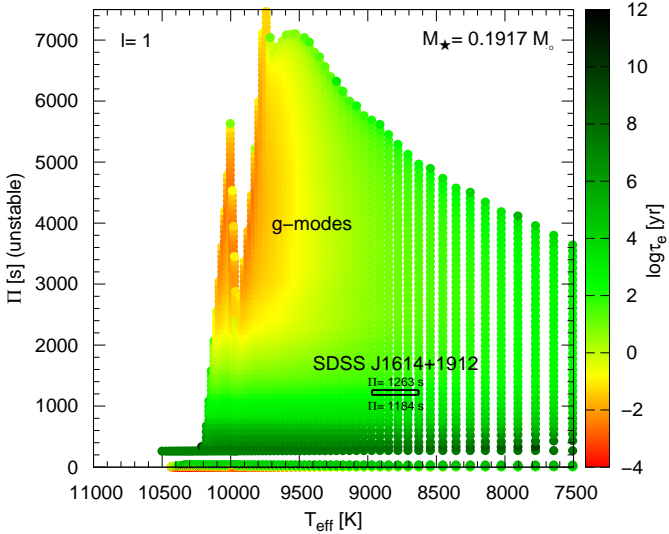} 
\caption{Similar to Fig. \ref{SDSSJ222859}, but for the the $0.1917 M_{\sun}$
sequence and the ELMV star SDSS J1614$+$1912.}
\label{SDSSJ161431} 
\end{center}
\end{figure}

The ELMV star SDSS J1614$+$1912 also was discovered to be pulsating by
\citet{2013MNRAS.436.3573H}. This star has $T_{\rm eff} \sim 8800$ K
and $M_{\star} \sim 0.19 M_{\sun}$, and pulsate  in just two periods,
at 1184.1 s and 1262.7 s. These are relatively short periods  when
compared with the periods detected in the other ELMVs, except SDSS
J1112+1117 (see below). This is something striking in view that, being
the second  coldest known ELMV star (after SDSS J2228$+$3623),  its
relatively short periods do not match the well known trend in ZZ Ceti
stars of an increase of pulsation periods for lower effective
temperatures
\citep{1993BaltA...2..407C,2006ApJ...640..956M}. According to its
stellar mass, we have to compare these periods with the theoretical
range of unstable mode periods of our sequence with $M_{\star}= 0.1917
M_{\sun}$. This comparison is displayed  in
Fig. \ref{SDSSJ161431}. Clearly, both periods are well accounted for
by the  theoretical computations.  

\begin{figure} 
\begin{center}
\includegraphics[clip,width=9 cm]{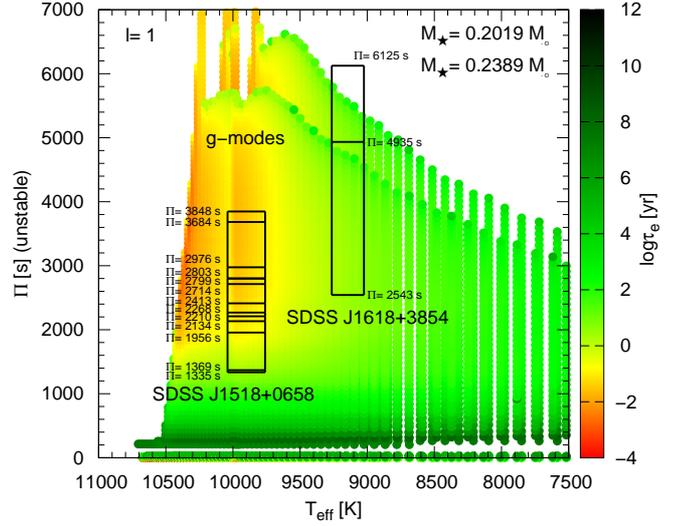} 
\caption{Similar to Fig. \ref{SDSSJ222859}, but for the the $0.2019 M_{\sun}$
and $0.2389 M_{\sun}$ sequences and the ELMV stars SDSS J1518$+$0658  
and SDSS J1618$+$3854.}
\label{SDSSJ151826andSDSSJ161831} 
\end{center}
\end{figure}

Next, we focus our attention on the ELMV stars SDSS J1518$+$0658
\citep{2013ApJ...765..102H} and  SDSS J1618$+$3854
\citep{2015ASPC..493..217B}, which have $T_{\rm eff} \sim 9900$ K and
$T_{\rm eff} \sim 9140 $ K , respectively.  With a stellar mass of
$M_{\star}\sim  0.22 M_{\sun}$ estimated for these  two stars, they
are the most massive ELMVs known hitherto. We compare the  observed
periods with the theoretical predictions corresponding to the
evolutionary sequences with $M_{\star}= 0.2019 M_{\sun}$ and
$M_{\star}= 0.2389 M_{\sun}$,  thus embracing the stellar mass derived
for both stars. We show the results in
Fig. \ref{SDSSJ151826andSDSSJ161831}. The 13 periods exhibited by SDSS
J1518$+$0658 in the range  $1335-3848$ s are well supported by our
nonadiabatic computations. Indeed, the detected periods, particularly
those longer than $\sim 2000$ s, correspond to the most unstable
theoretical $g$ modes of the instability domain, characterized by
$e$-folding times  in the range $10^{-2}-10^{-1}$ yrs. In the case of
SDSS J1618$+$3854, our  theoretical computations are successful in
reproducing the shortest observed periods at 2543 s and 4935 s, but
fail  in predict the existence of the longest one (6125 s).
So, if our nonadiabatic models are a good representatation 
of ELMVs, we can rule out the masses $0.2019 M_{\sun}$ and 
$0.2389 M_{\sun}$ for this star from the exhibited period range 
alone.

\begin{figure} 
\begin{center}
\includegraphics[clip,width=9 cm]{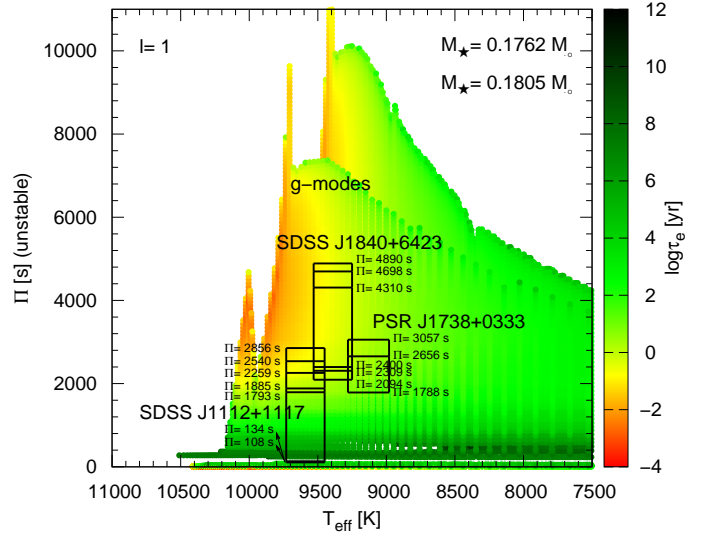} 
\caption{Similar to Fig. \ref{SDSSJ222859}, but for the the $0.1762 M_{\sun}$
and $0.1805 M_{\sun}$ sequences and the ELMV stars SDSS J1112+1117,
SDSS J1840+6423, and PSR J1738+0333.}
\label{SDSSJ111215_SDSSJ184037_PSRJ1738} 
\end{center}
\end{figure}

Finally, in Fig. \ref{SDSSJ111215_SDSSJ184037_PSRJ1738} we depict  the
situation for the remainder three ELMV stars, SDSS J1112+1117
\citep{2013ApJ...765..102H},  SDSS J1840+6423
\citep{2012ApJ...750L..28H}, and  PSR J1738+0333
\citep{2015MNRAS.446L..26K}. These stars have stellar masses
$M_{\star}/M_{\sun} \sim  0.179, 0.183$, and $0.181$, respectively,
near the critical mass for the development of CNO flashes  ($\sim 0.18
M_{\sun}$). As it was shown in \citet{2014A&A...569A.106C},  ELM stars
in this range of masses can harbor very different internal chemical
structures and in particular quite distinct H layer thicknesses, which
should be reflected in their pulsation spectra. So, future
asteroseismological  analysis  of these stars will have the potential
of place strong  constraints on the previous evolutionary history of
their progenitors.  We include in
Fig. \ref{SDSSJ111215_SDSSJ184037_PSRJ1738} the  domains of unstable
mode periods corresponding to the sequences with $M_{\star}= 0.1762
M_{\sun}$ and $M_{\star}= 0.1805 M_{\sun}$,  thus enclosing the masses
inferred for the three stars. The figure reveals that the periods
measured in these stars are well accounted  for by our stability
computations. The case of SDSS J1112+1117 is particularly interesting
because this is the only ELMV  star showing short periods (108 s and
134 s), in addition to  long periods ($\sim 1800-2900$ s) typical of
the class.   \citet{2013ApJ...765..102H} proposed the possibility that
these short  periods could be associated to $p$ modes. This idea was
examined  by \citet{2014A&A...569A.106C}, who found that, in the frame
of the  models of \citet{2013A&A...557A..19A}, if the temperature and
mass (gravity)  of the star are correct, the short periods cannot be
attributed  to $p$ modes nor radial modes. Alternatively,
\citet{2014ApJ...793L..17C} demonstrated that these short periods can
be associated to low-order $g$ modes destabilized mainly by the
$\varepsilon$-mechanism by stable nuclear burning at the basis of the
H envelope. 

\subsection{Using $T_{\rm eff}$ and $\log g$ corrected by 3D model atmosphere
effects} 
\label{3D}

Here, we assess how well our theoretical computations fit the
observations when we adopt the ELMV $T_{\rm eff}$ and $\log g$
parameters after 3D  corrections as given by
\citet{2015arXiv150701927T}.  Note that the situation for SDSS
J2228+3623 does not change, because  its effective temperature do not
experience a significant shift when 3D corrections are taken into
account ($\Delta T_{\rm eff}\sim 20$ K; see
Table\ref{tabla-ELMVs}) and,  although the $\log g$ shift is large, 
the mass of the star still is below to $0.1554 M_{\sun}$. 
Thus, the comparison already  shown in
Fig. \ref{SDSSJ222859} holds even when we adopt the 3D corrected
($T_{\rm eff}, \log g$) for this star. Since the long-period
boundary of the domain of unstable modes is longer for 
lower stellar masses, we conclude that the theoretical computations
are in good agreement with the range of excited periods
of SDSS J2228+3623. In the case of SDSS J1840+6423,
on the other hand, since the stellar mass is $\sim 0.177 M_{\sun}$
when we consider 3D corrected parameters  (see Fig. \ref{figure_02}), the
comparison between observed and theoretical ranges  of excited periods
shown in Fig. \ref{SDSSJ111215_SDSSJ184037_PSRJ1738} is still valid.
In particular, the observed range of excited periods is well accounted
for by  the theoretical computations corresponding to the sequence
with $M_{\star}= 0.1762 M_{\sun}$   when the effective temperature of
this star shifts from $T_{\rm eff}= 9390$ K (1D) to  $T_{\rm eff}=
9120$ K (3D). 

\begin{figure} 
\begin{center}
\includegraphics[clip,width=9 cm]{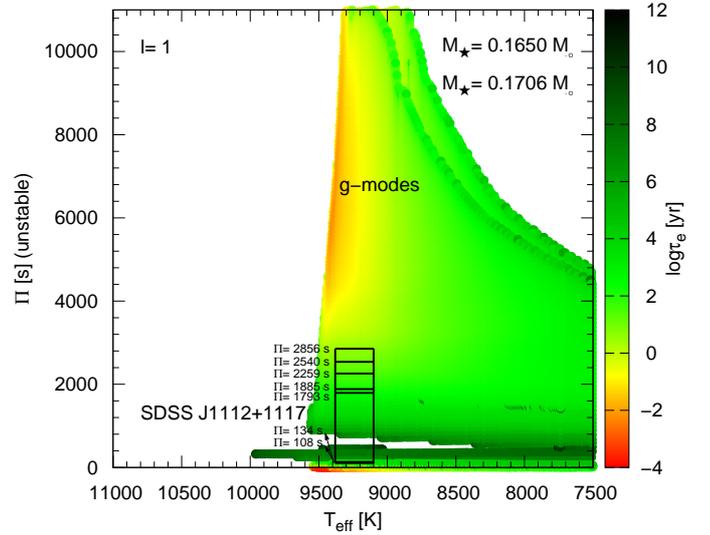} 
\caption{The periods of unstable $\ell= 1$ modes in terms of the
  effective temperature, corresponding to the $0.1650 M_{\sun}$ and
  $0.1706 M_{\sun}$ sequences. Also shown are the pulsation  periods
  of the ELMV star SDSS J1112+1117. The $T_{\rm eff}$ adopted for the
  star is its 1D value corrected by 3D model atmosphere effects (see
  Table\ref{tabla-ELMVs}).}
\label{SDSSJ111215} 
\end{center}
\end{figure}

For the remainder five ELMVs, the $T_{\rm eff}$, $\log g$ and
$M_{\star}$ change  substantially when we correct by 3D effects, and
we must make  further comparisons. To begin with, in
Fig. \ref{SDSSJ111215} we show the case of SDSS J1112$+$1117, in
which the observed periods  are compared with the excited theoretical
periods of the $0.1650 M_{\sun}$ and $0.1706 M_{\sun}$  sequences. In
this case, the $T_{\rm eff}$ turns out to be $350$ K lower  and the
stellar mass goes from $0.179 M_{\sun}$ to $0.169 M_{\sun}$, when we
take into account the 3D corrections. Clearly, the observed range of
periods is  well reproduced by  the theoretical
computations. Interestingly, we found that in this case  the shortest
periods at 108 s and 134 s could be safely identified with the $k= 1$
$p$ mode  of the $0.1650 M_{\sun}$ sequence, at variance with the
conclusion of \citet{2014A&A...569A.106C} who considered the $T_{\rm
  eff}$ and $\log g$ derived from 1D model atmospheres. 

\begin{figure} 
\begin{center}
\includegraphics[clip,width=9 cm]{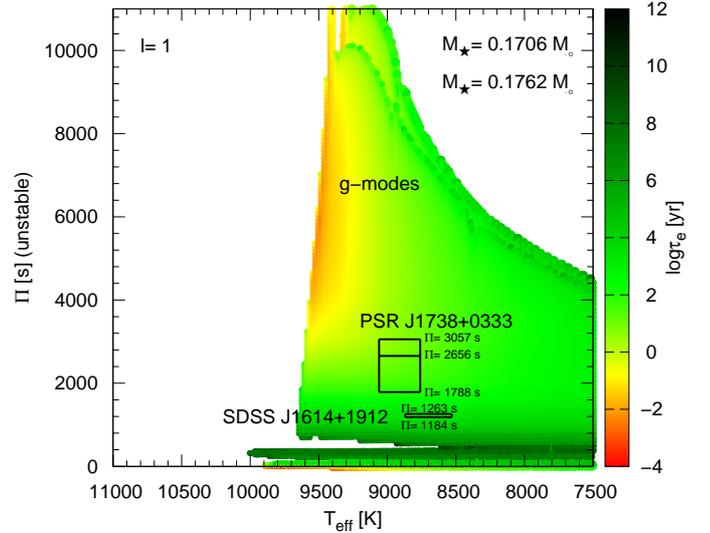} 
\caption{Similar to Fig. \ref{SDSSJ111215}, but for the case of PSR J1738+0333
and SDSS J1614+1912, and the  $0.1706 M_{\sun}$ and $0.1762 M_{\sun}$ sequences.}
\label{PSRJ1738_SDSSJ161431} 
\end{center}
\end{figure}

In Fig. \ref{PSRJ1738_SDSSJ161431} we illustrate the cases of PSR
J1738+0333 and SDSS J1614+1912,  where the observed ranges of periods
are compared with the theoretical ones  corresponding to the sequences
with $M_{\star}= 0.1706 M_{\sun}$ and  $M_{\star}= 0.1762 M_{\sun}$.
Clearly, the observed ranges of excited periods in these stars is well
accounted for the theoretical computations. 

\begin{figure} 
\begin{center}
\includegraphics[clip,width=9 cm]{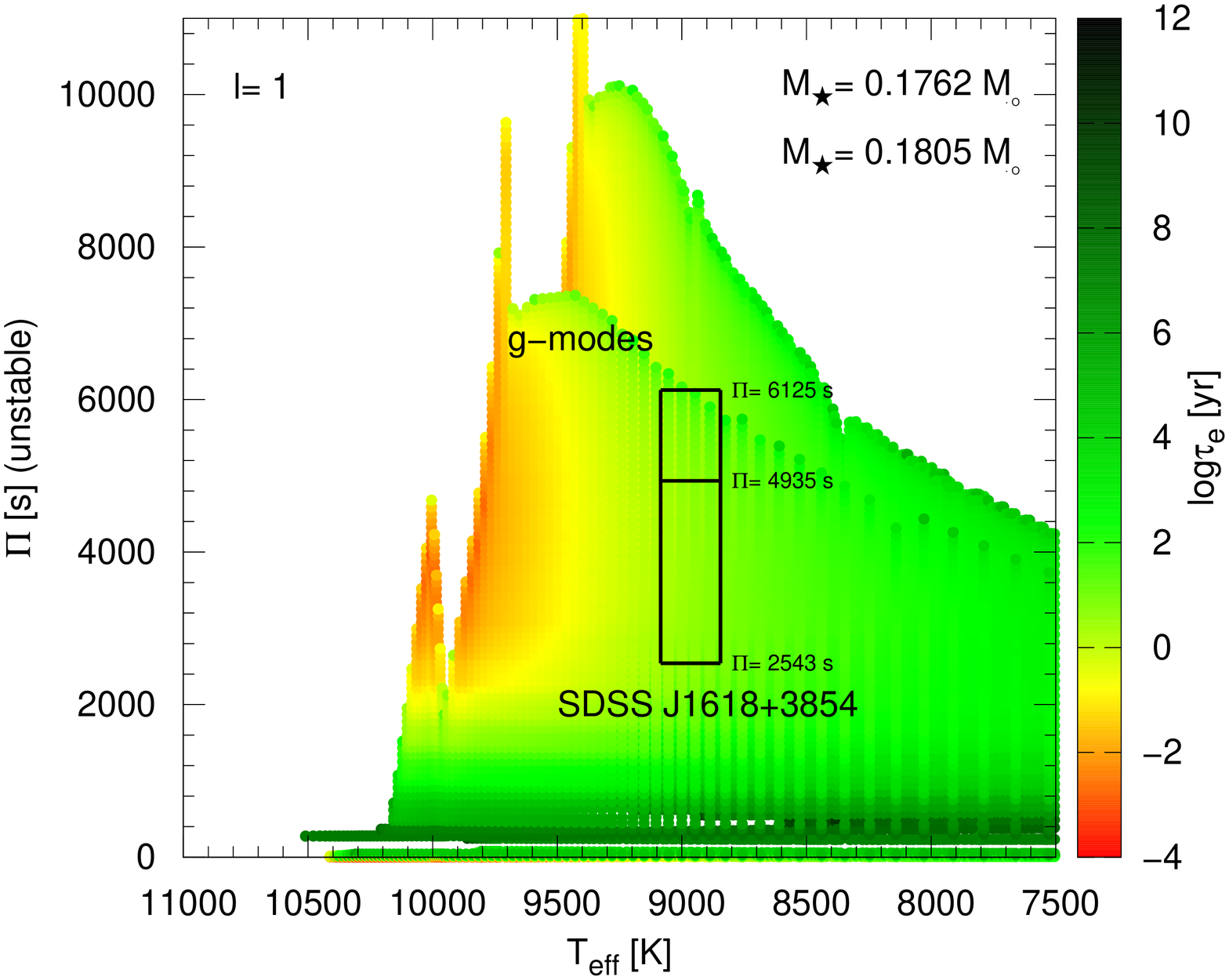} 
\caption{Similar to Fig. \ref{SDSSJ111215}, but for the case of SDSS
J1618+3854, and the  $0.1762 M_{\sun}$ and $0.1805 M_{\sun}$ sequences.}
\label{SDSSJ161831} 
\end{center}
\end{figure}

The situation for SDSS J1618+3854 is depicted in
Fig. \ref{SDSSJ161831}. The comparison  in this case is made with the
theoretical unstable modes corresponding to the $0.1762 M_{\sun}$ and
$0.1805 M_{\sun}$  sequences. There is a good agreement  between the
observations and the theoretical predictions. In particular,  the
existence of the longest period at 6125 s is reliably predicted by our
stability  computations. This is at variance with the case in which we
adopted the  $T_{\rm eff}$ and $\log g$ derived from 1D model
atmospheres  (see Fig. \ref{SDSSJ151826andSDSSJ161831}). 
This finding gives strong support to the 3D model atmosphere calculations of 
\citet{2015arXiv150701927T}.

\begin{figure} 
\begin{center}
\includegraphics[clip,width=9 cm]{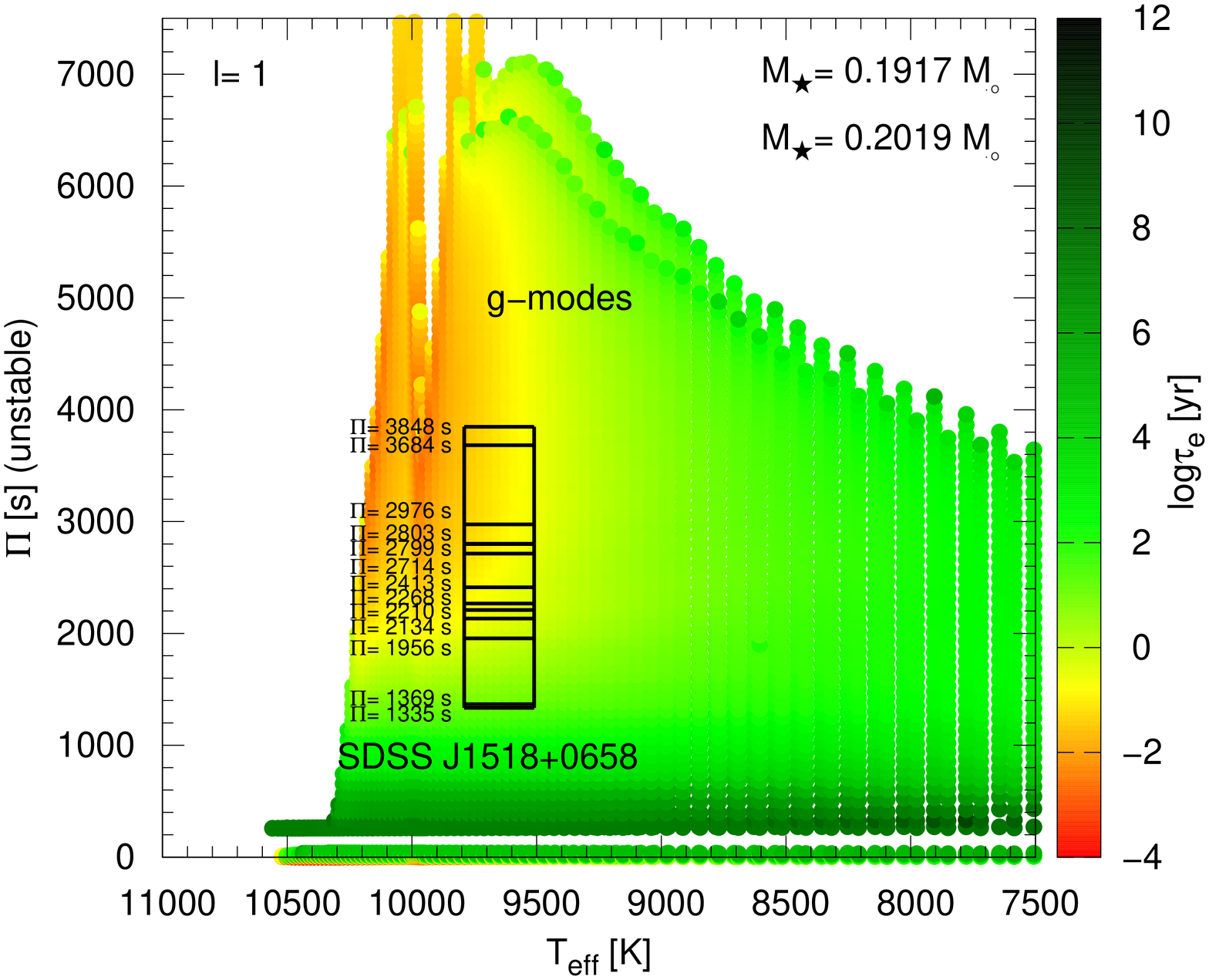} 
\caption{Similar to Fig. \ref{SDSSJ111215}, but for the case of SDSS J1518+0658, 
and the  $0.1917 M_{\sun}$ and $0.2019 M_{\sun}$ sequences.}
\label{SDSSJ151826} 
\end{center}
\end{figure}

Finally, we display in Fig. \ref{SDSSJ151826} the case of SDSS
J1518+0658, for which we  compare the observed range of periods with
the theoretical computations corresponding to the sequences with
$M_{\star}= 0.1917 M_{\sun}$ and $M_{\star}= 0.2019 M_{\sun}$. Again
in this  case our theoretical predictions are in excellent agreement
with the range of excited periods observed in the star.

We close this section by noting that for all the analyzed ELMV stars,
the number of periods detected is disappointingly low in comparison
with the rich  spectrum of periods of unstable modes, that include
radial and nonradial $p$ and $g$ modes,  as predicted by theoretical
computations. As for the other classes of  pulsating WDs, there must
be some unknown filter mechanism present in real stars,  that favors
only a few  periods (out of the available dense spectrum of
eigenmodes)  to reach observable amplitudes. Finding that missing
piece of physics in our pulsation models  is well beyond the scope of
the present paper.

\section{Summary and conclusions}
\label{conclusions}

In this paper, we have presented a detailed pulsation stability study
of  pulsating low-mass WDs employing the set of state-of-the-art
evolutionary models of \citet{2013A&A...557A..19A}.  This is the
second paper of a series on this topic,  being the first one focused
on the adiabatic properties of low-mass WDs
\citep{2014A&A...569A.106C}. Preliminary results of  the nonadiabatic
analysis detailed here have been already presented in
\citet{2014ApJ...793L..17C}, focused mainly on the role that stable H
burning has in destabilizing low-order $g$ modes of ELM  WD models.
In the present paper, we extend that analysis  by assessing the
pulsational stability of radial ($\ell= 0$)  and nonradial ($\ell= 1,
2$) $g$ and $p$ modes for the complete set  of 14 evolutionary
sequences of low-mass He-core WD models of
\citet{2013A&A...557A..19A} with masses in the range $0.1554-0.4352
M_{\sun}$, considering both the $\kappa-\gamma$ and $\varepsilon$
mechanisms of mode excitation, and including different prescriptions
of the MLT theory of convection.

Our main findings are:

\begin{itemize}

\item For all the model sequences analyzed, a dense spectrum of
  unstable  radial modes and nonradial $g$ and $p$ modes are excited
  by the $\kappa-\gamma$  mechanism due to the H partial ionization 
  zone in the stellar envelope.  In addition, some short-period $g$ modes
  are destabilized mainly  by the $\varepsilon$ mechanism due to
  stable nuclear burning at the basis of the  H envelope
  (Fig. \ref{figure_06}), particularly for model  sequences
  with $M_{\star} \lesssim 0.18 M_{\sun}$ (see Table
  \ref{table-epsilon}).

\item The blue edge of the instability domain in the $T_{\rm eff}-\log
  g$ plane is hotter for higher the stellar mass and convective
  efficiency   (Fig. \ref{figure_07}).  The ML2 and ML3 versions  
  of the
  MLT theory of convection are the only ones that correctly accounts for
  the location of the seven known ELMV stars, regardless of whether
  the $T_{\rm eff}$ and $\log g$ for the stars is derived from
  standard 1D model  atmospheres or if these parameters are corrected
  by 3D effects  (see Fig. \ref{figure_08}). There is no
  dependence of the blue edge  of $p$ modes with the harmonic degree;
  in the case of $g$ modes, we found a weak sensitivity of the blue
  edge with $\ell$. Finally,  the blue edges corresponding to radial
  and nonradial $p$ modes are  somewhat ($\sim 200$ K)  hotter  than
  the blue edges of $g$ modes.

\item Generally, the most unstable modes (shorter $e$-folding times)
  are those characterized by high and intermediate radial orders. For
  instance, in the case of the sequence with $M_{\star}=
  0.1762M_{\sun}$ and ML2, the most unstable modes have periods
  between $\sim 2000$ s and $\sim 10\,000$ s  ($k$ between  $\sim 20$
  and $\sim 110$) for $g$ modes, and periods between  $\sim 7$ s ($k=
  35$) and $\sim 30$ s ($k= 7$) for $p$ modes and radial modes. The
  most unstable modes correspond to stellar models  located near to
  the blue edge of the instability domain (see
  Fig. \ref{figure_09}).

\item The longest and shorter excited periods of $g$-modes are longer
  for lower $M_{\star}$ and smaller $\ell$. In the case of $p$ modes
  and radial modes, the longest and shortest excited periods are
  longer for  lower $M_{\star}$, although they do not depend on
  $\ell$.

\item For sequences with $M_{\star} \leq 0.1762 M_{\sun}$,  the
  longest excited periods of $g$ modes are substantially longer for
  higher convective efficiency. On the contrary, for  $M_{\star} \geq
  0.1805 M_{\sun}$ the longest excited periods of $g$ modes are
  shorter for higher convective efficiency,  although the differences
  are small. In the case of $p$ modes and radial modes,  we found a
  very weak trend of longer shortest and longest unstable periods with
  higher convective efficiency.

\item We compared the ranges of unstable mode periods predicted by our
  stability  analysis with the ranges of periods observed in the ELMV
  stars. Irrespective of whether  we adopt the ($T_{\rm eff}, \log g$)
  derived from 1D model atmospheres or  these parameters corrected by
  3D effects, we found generally an excellent agreement, as shown by
  Figs. \ref{SDSSJ222859} to \ref{SDSSJ151826}. 

\item In the specific case of SDSS J1618+3854, if we adopt $T_{\rm
  eff}$  and $\log g$ as derived from 1D model atmosphere
  computations, our  nonadiabatic models are unable to explain the
  existence of the longest period at 6125 s
  (Fig. \ref{SDSSJ151826andSDSSJ161831}).  However, this period is reliably
  predicted  when we adopt $T_{\rm eff}$ and $\log g$ values corrected
  by 3D model atmosphere effects (Fig. \ref{SDSSJ161831}).  This gives
  strong support to the 3D model atmospheres of
  \citet{2015arXiv150701927T}.

\end{itemize}

The results of this study, along with those of previous  research
\citep{2010ApJ...718..441S,
  2012A&A...547A..96C,2013ApJ...762...57V,2014ApJ...793L..17C}, allow
us to know the origin and basic nature of the pulsations exhibited  by
ELMV stars. However, even though theoretical models reproduce
qualitatively the observations, some essential unknowns still
remain. For instance, there is the problem of the red edge of the
instability strip. Our calculations,  which assume the FC
approximation, as well as those of \citet{2013ApJ...762...57V}, which
include a TDC treatment, predict a red edge extremely cool  ($T_{\rm
  eff}\sim 5000-6000$ K) as compared with the coolest ELMV star (SDSS
J2228+3623, $T_{\rm eff} \sim 7900$ K).  
Fortunately, this incomplete knowledge of the physics of WD pulsations
does not prevent us from moving forward in asteroseismological
studies based on \emph{adiabatic} calculations, in which it does not
matter the physical agent that drives  the pulsations, but the value
of the periods themselves, which depend sensitively on the internal
structure of the WD star.  Asteroseismological analysis will provide
valuable clues about the internal  structure and evolutionary status
of low-mass WDs, allowing us to place  constraints on the binary
evolutionary processes involved in their formation. But in order to
extend the parameter space to explore,  we have to consider a possible
range of H envelope thicknesses.  We plan to compute new evolutionary
sequences of low-mass He-core WDs with different angular-momentum loss
prescriptions due to mass loss, which could have an impact on the
final H envelope mass. Results of these investigations will be
presented in an upcoming paper.

\begin{acknowledgements}
We wish to thank our anonymous referee for the constructive
comments and suggestions that greatly improved the original version of
the paper. We warmly thank K. J. Bell and J. J. Hermes for reading the paper 
and making enlightening comments and suggestions.  
Part of this work was supported by AGENCIA through the Programa de
Modernizaci\'on Tecnol\'ogica BID 1728/OC-AR, and by the PIP
112-200801-00940 grant from CONICET. This research has made use of
NASA Astrophysics Data System.
\end{acknowledgements}


\bibliographystyle{aa} 
\bibliography{paper} 

\end{document}